\documentclass{elsart}
\usepackage[dvips]{graphics}
\usepackage{graphicx}
\usepackage{amsmath}
\usepackage{amssymb}
\usepackage{amsfonts}

\journal{Annals of Physics}

\begin{document}

\begin{frontmatter}

\title{Quantum fields in toroidal
topology}

\author[UA,TRIUMF]{F.C. Khanna},
\ead{khanna@phys.ualberta.ca}
\author[CBPF]{A.P.C. Malbouisson},
\ead{adolfo@cbpf.br}
\author[UFBA]{J.M.C. Malbouisson},
\ead{jmalboui@ufba.br}
\author[UA,UNB]{A.E. Santana\corauthref{cor1}}
\ead{asantana@unb.br}

\corauth[cor1]{Corresponding author}

\address[UA]{Theoretical Physics Institute, University of Alberta,
Edmonton, AB T6G 2J1, Canada}
\address[TRIUMF]{TRIUMF, Vancouver, BC, V6T 2A3, Canada}
\address[CBPF]{Centro Brasileiro de Pesquisas F\'{\i}sicas/MCT,
22290-180, Rio de Janeiro, RJ, Brazil}
\address[UFBA]{Instituto de F\'{\i}sica, Universidade Federal da
Bahia, 40210-340, Salvador, BA, Brazil}
\address[UNB]{Instituto de F\'{\i}sica, International Center for Condensed
Matter Physics, Universidade de Bras\'{\i}lia, 70910-900,
Bras\'{\i}lia, DF, Brazil}

\begin{abstract}
The standard representation of c*-algebra is used to describe fields
in compactified space-time dimensions characterized by topologies of
the type $ \Gamma _{D}^{d}=(\mathbb{S}^{1})^{d}\times
\mathbb{M}^{D-d}$. The modular operator is generalized to introduce
representations of isometry groups. The Poincar\'{e} symmetry is
analyzed and then we construct the modular representation by using
linear transformations in the field modes, similar to the Bogoliubov
transformation. This provides a mechanism for compactification of
the Minkowski space-time, that follows as a generalization of the
Fourier-integral representation of the propagator at finite
temperature. An important result is that the $2\times2$ representation of the
real time formalism is not needed. The end result on calculating
observables is described as a condensate in the ground state. We
analyze initially the free Klein-Gordon and Dirac fields, and then
formulate non-abelian gauge theories in $\Gamma _{D}^{d}$. Using the
S-matrix, the decay of particles is calculated in order to show the
effect of the compactification.
\end{abstract}

\end{frontmatter}

\section{Introduction}

The thermal quantum field theory was proposed by Matsubara~\cite{3mats1},
based on an imaginary-time defined by a Wick rotation of the time-axis. The
propagator is written as a Fourier series in the imaginary time, by using
the frequencies: $\omega _{n}=2\pi n/\beta $ ($\pi (2n+1)/\beta $) for
bosons (fermions), corresponding to the period $\beta =T^{-1},$ with $T$
being the temperature~\cite{3ume4,3kap1}. Using the notion of spectral
function, as discussed by Kadanof and Baym~\cite{baym11}, Dolan and Jackiw~%
\cite{3dj} managed to write the thermal propagator in a Fourier integral
representation, thus restoring then time as a real quantity. In other words,
the imaginary-time formalism is interpreted topologically. It is proved that
the temperature is introduced in a quantum field theory by writing the
original theory, formulated in the Minkowski space, $\mathbb{M} ^{4}$, in
the compactified manifold $\Gamma _{4}^{1}=\mathbb{S}^{1}\times \mathbb{M}%
^{3},$ where the compactified dimension is the imaginary time. The
circumference of $\mathbb{S}^{1}$ is $\beta $~\cite{birrel1,kha1}. It is our
objective here to generalize the argument advanced in Refs.~\cite{baym11} and \cite{3dj} to
include not only the time but also space coordinates, in such way that any set
of dimensions of the manifold $\mathbb{M}^{D}$ can be compactified, defining
a theory in the topology $\Gamma _{D}^{d}=(\mathbb{S} ^{1})^{d}\times
\mathbb{M}^{D-d}$, with $1\leq d\leq D$. This establishes that the Fourier
integral representation is sufficient to deal with the general question of
compactification in the $\Gamma _{D}^{d}$ topology at finite-temperature and
real-time. To proceed, we have to set forth the general structure for such
an approach. As a consequence, for the case of $\Gamma _{4}^{1}$, that
structure provides a simplified version of the real-time formalism for
finite-temperature quantum fields.

It is well-known that there are two versions for a real-time
finite-temperature quantum field theory. One was formulated by Schwinger~%
\cite{3kel1,3kel2,3kel3} and Keldysh~\cite{3kel4}, and is based on using a
path in the complex time plane~\cite{3belac1}. The other is the thermofield
dynamics (TFD) proposed by Takahashi and Umezawa~\cite{3ume1}. In this case,
the thermal theory is constructed on a Hilbert space and thermal effects are
introduced by a Bogoliubov transformation~\cite{kha1,3ume1,3ume2,3das}. In
equilibrium, these two real-time formalisms are basically the same and the
propagator, $G^{ab},\ a,b=1,2$, is a $2\times 2$ matrix~\cite{chu1}.
However, the physical content is present only in the component $G^{11}$,
which is, moreover, just the propagator in the Fourier integral
representation as studied by Dolan and Jackiw. Such a result suggests that a
real-time theory may be fully formulated in such a way that the propagator
is written as a $c$-number (not as a matrix). This is an additional
motivation for developing the central theme of this paper, insofar as this
procedure can be extended to general cases with the topology $\Gamma _{D}^{d}$.

An approach describing systems in compactified spaces is derived as a
generalization of both the Matsubara formalism, involving the Fourier
series,~\cite{birrel1,birrel2,3comp1,3comp2} and TFD~\cite{3comp3}. There
are numerous applications of such a formalism, including the Casimir effect
for the electromagnetic and fermion fields within a box~\cite{3comp3,3comp4}%
, the $\lambda \phi ^{4}$ model describing the order parameter for the
Ginsburg-Landau theory for superconductors~\cite{3comp5}, and the
Gross-Neveu model as an effective approach for QCD~\cite{3comp6,KMMS10}. The
extension of this method to the Fourier integral representation is important
to address many other problems in a topology $\Gamma _{D}^{d}$ that are of
interest in different areas, such as cosmology, condensed matter and
particle physics~\cite%
{ford3,genr8,chodos22,tori1,eliz1,gen6,ito1,gen3,genr9,gen7,gen1,luc1,luc2,luc3,mul1,genr4,ebe1,ebe2,ebe5}.
In order to proceed with such a generalization, we rely on algebraic
bases, using the modular representation of the $c^{\ast }$-algebra.

The $c^{\ast }$-algebra has played a central role in the development of
functional analysis, and has attracted much attention due to its importance in
non-commutative geometry~\cite{aco1,aco2,ebe3,kraj1}. It was first
associated with the quantum field theory at finite temperature, through the
imaginary time formalism~\cite{emch}. Actually, the search for the algebraic
structure of the Gibbs ensemble theory leads to the Tomita-Takesaki, or the
standard representation of the $c^{\ast }$-algebra~\cite{emch,brate,tomita}.

For the real-time formalism, the $c^{\ast }$-algebra approach was first
analyzed by Ojima~\cite{ojima1}. Later, the use of the Tomita-Takesaki
Hilbert space, as the carrier space for representations of kinematical
groups, was developed~\cite{ade31,ade44}, in order to build thermal theories
from symmetry: that is called the thermo group. A result emerging from this
analysis is that the tilde-conjugation rules, the doubling in TFD, are
identified with the modular conjugation of the standard representation. In
addition, the Bogoliubov transformation, the other basic TFD ingredient,
corresponds to a linear mapping involving the commutants of the von Neumann
algebra. This TFD apparatus has been developed and applied in a wide range
of problems~\cite{3ume2,3das,kha1000}. In particular, the physical
interpretation of the doubling of the operators has been identified~\cite%
{kha1}.

The doubling of the propagator in real-time formalisms, however, is no
longer necessary since we consider, as a starting point, the modular group
introduced in a $c^{\ast }$-algebra. This is a result that we prove here, by
using the modular representation of the Poincar\'{e} group. It is worth
mentioning that, the modular conjugation is defined in order to respect the
Lie algebra structure. This procedure provides a consistent way to define
the modular conjugation for fermions; which is usually a non-simple task due
to a lack of criterion~\cite{ojima1}. Using a Bogoliubov transformation, the
effect of compactification in $\Gamma _{D}^{d}$ is introduced and a Fourier
integral representation for the propagator is derived. Initially, an
analysis is carried out for free boson and fermion fields. An extension of
the formalism for abelian and non-abelian gauge-fields is introduced by
functional methods. Exploring the canonical formalism, the S-matrix is
developed and applied to calculate, as an example, decay rates by
considering compactified spatial dimensions. It is important to emphasize
that, the Feynman rules follow in parallel as in the Minkowski space-time
theory and that the compactification corresponds to a process of
condensation in the vacuum state. The topology then leads naturally to the
notion of quasi-particles.

The paper is organized in the following way. In Section 2, we present
elements of the standard representation of the $c^{\ast }$-algebra, and use it to
construct a representation theory of quantum fields in $\Gamma _{D}^{d}$
topologies. In Section 3, the modular representations of Lie algebras are
developed. In Section 4, physical aspects of the theory are discussed. In
Section 5, we construct the path integral for determining fields in
compactified space-time. In section 6, we elaborate the extension of the
formalism to an $SU(3)$ gauge theory. In Section 7, the $S$-matrix is
developed with application to the analysis of reaction rates. Concluding
remarks are presented in Section 8.

\section{ C$^{\ast }$-algebra and compactified propagators}

Let us initially present a r\'{e}sum\'{e} of some aspects of
$c^{\ast }$-algebras in order to \ explore representations of Lie
groups~\cite{ade31,ade44}. This sets the basis to built up the
appropriate generalization of the finite-temperature formalism to
also accommodate spatial compactification.

A $c^{\ast }$-algebra $\mathcal{A}$ is a von Neumann algebra
over the field of complex numbers $\mathbb{C}$ with two different maps, an
involutive mapping $^{\ast }:\mathcal{A}\rightarrow \mathcal{A\ }$ and the
\textit{norm}, which is a mapping defined by $\Vert \cdot \Vert :\mathcal{A}%
\rightarrow \mathbb{R}\mathbf{_{+\text{ }}}$~\cite{emch,brate,tomita}. Let $(%
\mathcal{H}_{w},\pi _{w}(\mathcal{A}))$ be a faithful realization of $%
\mathcal{A}$, where $\mathcal{H}_{w}$ is a Hilbert space and $\pi _{w}(%
\mathcal{A}):\mathcal{H}_{w}\rightarrow \mathcal{H}_{w}$ is a $^{\ast }$%
-isomorphism of $\mathcal{A}$ defined by linear operators in $\mathcal{H}%
_{w} $. Taking $\mid \xi _{w}\rangle \in \mathcal{H}_{w}$ to be normalized,
it follows that $\langle \xi _{w}\mid \pi _{w}(A)\mid \xi _{w}\rangle ,$ for
every $A\in \mathcal{A}$, defines a state over $\mathcal{A}$ denoted by $%
w_{\xi }(A)=\langle \xi _{w}\mid \pi _{w}(A)\mid \xi _{w}\rangle $. As was
demonstrated by Gel'fand, Naimark and Segal (GNS), the inverse is also true;
i.e. every state $\omega $ of a $c^{\ast }$-algebra $\mathcal{A}$ admits a
vector representation $\mid \xi _{w}\rangle \in \mathcal{H}_{w}$ such that $%
w(A)\equiv \langle \xi _{w}\mid \pi _{w}(A)\mid \xi _{w}\rangle $. This
realization is called the GNS construction~\cite{emch,brate,tomita}, which
is valid if the dual coincides with the pre-dual.

The Tomita-Takesaki (standard) representation is a class of representations
of $c^{\ast }$-algebras introduced as follows. Consider $\sigma :\mathcal{H}
_{w}\rightarrow \mathcal{H}_{w}$ to be a (modular) conjugation in $\mathcal{%
H }_{w}$, that is, $\sigma $ is an anti-linear isometry such that $\sigma
^{2}=1$. The set $(\mathcal{H}_{w},\pi _{w}(\mathcal{A}))$ is a
Tomita-Takesaki representation of $\mathcal{A}$, if $\sigma \pi _{w}(
\mathcal{A})\sigma =\widetilde{\pi }_{w}(\mathcal{A})$ defines a $^{\ast }$
-anti-isomorphism on the linear operators. It follows that $(\mathcal{H}%
_{w}, \widetilde{\pi }_{w}(\mathcal{A}))$ is a faithful anti-realization of $%
\mathcal{A}$. It is to be noted that $\widetilde{\pi }_{w}(\mathcal{A})$ is
the commutant of $\pi _{w}(\mathcal{A})$; i.,e. $[\pi _{w}(\mathcal{A}),
\widetilde{\pi }_{w}(\mathcal{A})]=0$. In this representation, the state
vectors are invariant under $\sigma $; that is, $\sigma \mid \xi _{w}\rangle
=\mid \xi _{w}\rangle $. As long as there is no confusion, elements of the
set $\pi _{w}(\mathcal{A})$ will be denoted by $A$ and those of $\widetilde{%
\pi } _{w}(\mathcal{A})$ by $\widetilde{A}$.

With this notation, the tilde and non-tilde operators, defined above by the $%
\sigma $ modular conjugation, have the properties,
\begin{align*}
(A_{i}A_{j})\ ^{\widetilde{}}& =\widetilde{A}_{i}\widetilde{A}_{j}, \\
(cA_{i}+A_{j})\ ^{\widetilde{}}& =c^{\ast }\widetilde{A}_{i}+\widetilde{A}
_{j}, \\
(A_{i}^{\dagger })\ ^{\widetilde{}}& =(\widetilde{A}_{i})^{\dagger }, \\
\widetilde{\widetilde{A}}_{i}& =A_{i}, \\
\lbrack A_{i},\widetilde{A}_{j}]& =0, \\
|\xi _{w}\rangle ^{\widetilde{}\ }& =|\xi _{w}\rangle \,\,\text{%
and\thinspace \thinspace }\langle \xi _{w}|^{\widetilde{}}=\langle \xi _{w}|.
\end{align*}
These tilde-conjugation rules are derived in TFD in association with properties of physical
(usually non-interacting) systems and are called the tilde conjugation rules~\cite{3ume2}.

An interesting aspect of this construction is that properties of $^{\ast }$-automorphisms
in $\mathcal{A}$ can be defined through a unitary operator,
say $\Delta (\tau )$, invariant under the modular conjugation, i.e. $[\Delta
(\tau ),\sigma ]=0$. Then writing $\Delta (\tau )$ as $\Delta (\tau )=\exp
(i\tau \widehat{A})$, where $\widehat{A}$ is the generator of symmetry, we
have $\sigma \widehat{A}\sigma =-\widehat{A}$. Therefore, the generator $%
\widehat{A}$ is an odd polynomial function of $A-\widetilde{A}$, i.e.
\begin{equation}
\widehat{A}=f(A-\widetilde{A})=\sum_{n=0}^{\infty }c_{n}(A-\widetilde{A}
)^{2n+1},  \label{um}
\end{equation}%
where the coefficients $c_{n}\in \mathbb{R}$.

Consider the simple case where $c_{0}=1,$ $c_{n}=0,\forall \ n\neq 0,$ i.e. $%
\widehat{A}=A-\widetilde{A}.$ Taking $A$ to be the Hamiltonian, $H,$ the
time-translation generator is given by $\widehat{H}$ . The parameter $\tau $
is related to a Wick rotation such that $\tau \rightarrow i\beta $;
resulting in $\Delta (\beta )=e^{-\beta \widehat{H}}$, where $\beta =T^{-1}$
, $T$ being the temperature. This is the so-called modular operator in $%
c^{\ast }$-algebras. As a consequence, a realization for $w(A)$ as a Gibbs
ensemble average is introduced~\cite{emch,brate,tomita},
\begin{equation}
w_{A}^{\beta}=\frac{\mathrm{Tr}(e^{-\beta H}A)}{\mathrm{Tr}e^{-\beta H}}.
\label{ade11}
\end{equation}

We now proceed with the generalization of this construction for
finite temperature, corresponding to the compactification of the
imaginary time, to accommodate also spatial compactification. To do
so, we replace $H$ by the generator of space-time translations,
$P^{\mu }$, in a $d$-dimensional subspace of a $D$-dimensional
Minkowski space-time, $\mathbb{M}^{D}$, with $d\leq D$. Then we
generalize Eq.~(\ref{ade11}) to the form
\begin{equation}
w_{A}^{\alpha }=\frac{\mathrm{Tr}(e^{-\alpha _{\mu }P^{\mu }}A)}{\mathrm{Tr}
e^{-\alpha _{\mu }P^{\mu }}},  \label{adol11}
\end{equation}
where $\alpha _{\mu }$ are the group parameters. This leads to the following
statement:

\begin{itemize}
\item Proposition 1: For $A(x)$ in the $c^{\ast}$-algebra $\mathcal{A}$,
there is a function $w_{A}^{\alpha }(x)$, $x$ in
$\mathbb{M}^{D}\mathbf{,}$ defined by Eq.~(\ref{adol11}) such that
\begin{equation}
w_{A}^{\alpha }(x)=w_{A}^{\alpha }(x+i\alpha ),  \label{adol13}
\end{equation}
where $\alpha =(\alpha _{0},\alpha _{1},...,\alpha _{d-1},0,\dots,0)$. This implies
that $w_{A}^{\alpha }(x)$ is periodic, in the $d$-dimensional subspace, with
$\alpha _{0}$ being the period in the imaginary time, i.e. $\alpha
_{0}=\beta $, and $\alpha _{j} = i L_{j}$, $j=1,...,d-1$, are identified
with the periodicity in spatial coordinates.
\end{itemize}

It is important to be noted that $w_{A}^{\alpha }(x)$ preserves the isometry, since it
is defined by elements of the isometry group. Therefore, the theory is
defined in the topology $\Gamma _{D}^{d}=(\mathbb{S}^{1})^{d}\times \mathbb{M%
}^{D-d}$. For the particular case of $d=1$, taking $\alpha _{0}=\beta $, we
have to identify Eq.~(\ref{adol13}) as the KMS (Kubo-Martin-Schwinger)
condition~\cite{kha1}. Then by using the GNS construction, a quantum theory
in thermal equilibrium is equivalent to taking this theory in a $\Gamma
_{D}^{1}$ topology in the imaginary-time axis, where the circumference of $%
\mathbb{S}^{1}$ is $\beta $, as it is well known. The generalization of this
result for space coordinates is given by Eq.~(\ref{adol13}); it corresponds
to the generalized KMS condition for field theories in toroidal spaces $(%
\mathbb{S}^{1})^{d}\times \mathbb{M}^{D-d}$. On the other hand the average
given in Eq.~(\ref{adol11}) can also be written as%
\begin{equation}
w_{A}^{\alpha }(x)\equiv \langle \xi _{w}^{\alpha }|A(x)|\xi_{w}^{\alpha} \rangle .  \label{CBPF1}
\end{equation}%
In the next section, we turn our attention to constructing the state $|\xi_{w}^{\alpha} \rangle$ explicitly.

Let us consider, as an example, the free propagator for the
Klein-Gordon-field. In this case the Lagrangian density is
\begin{equation*}
\mathcal{L}=\frac{1}{2}\partial _{\mu }\phi (x)\partial ^{\mu }\phi (x)-%
\frac{m^{2}}{2}\phi (x)^{2},
\end{equation*}%
and we take
\begin{equation*}
A(x,x^{\prime })=\mathrm{T}[\phi (x)\phi (x^{\prime })],
\end{equation*}%
where $\mathrm{T}$ is the time-ordering operator. The propagator for the
compactified field is $G_{0}(x-x^{\prime };\alpha )\equiv w_{A}^{\alpha
}(x,x^{\prime })$, that is
\begin{equation}
G_{0}(x-x^{\prime };\alpha )=\frac{\mathrm{Tr}(e^{-\alpha _{\mu }P^{\mu
}}T\phi (x)\phi (x^{\prime }))}{\mathrm{Tr}e^{-\alpha _{\mu }p^{\mu }}}\,.
\label{aug09148}
\end{equation}%
The generalized KMS condition, Eq.~(\ref{adol13}), then allows us to write
the series-integral representation, corresponding to a modified Matsubara
prescription, i.e.
\begin{equation}
G_{0}(x-x^{\prime };\alpha )=\frac{1}{i^{d}\,\alpha _{0}\cdot \cdot \cdot
\alpha _{d-1}}\sum\limits_{n_{0}\cdot \cdot \cdot n_{d-1}}\int \frac{d^{D-d}%
\mathbf{k}}{(2\pi )^{D-d}}\,\frac{e^{-ik_{\alpha }\cdot (x-x^{\prime })}}{%
k_{\alpha }^{2}-m^{2}+i\varepsilon },  \label{may10201}
\end{equation}%
where $k_{\alpha }=(k_{n_{0}}^{0},k_{n_{1}}^{1},\dots
,k_{n_{d-1}}^{d-1},k^{d}\dots ,k^{D-1})$, with
\begin{equation*}
k_{n_{j}}^{j}=\frac{2\pi n_{j}}{\alpha _{j}},\ \ 0\leq j\leq d-1,
\end{equation*}%
$n_{j}\in \mathbb{Z}$ and $d^{D-d}\mathbf{k}=dk^{d}dk^{d+1}\cdots dk^{D-1}.$
In what follows, we employ the name Matsubara representation (or
prescription) referring to both time and space compactification. The Green
function $G_{0}(x-x^{\prime };\alpha )$ is a solution of the Klein-Gordon
equation since $w_{A}^{\alpha }$, by definition, respects the isometry. This
means that $G_{0}(x-x^{\prime };\alpha )$ is the Green function of a boson
field defined locally in the Minskowki space-time. Globally this theory is
such that $G_{0}(x-x^{\prime };\alpha )$ has to satisfy periodic boundary
conditions. These facts assure us that $G_{0}(x-x^{\prime };\alpha )$ \ is
the Green function of a field theory defined in a hyper-torus, $\Gamma
_{D}^{d}=(\mathbb{S}^{1})^{d}\times \mathbb{M}^{D-d}$, with $1\leq d\leq D$,
where the circumference of the $j$-th $\mathbb{S}^{1}$ is specified by $%
\alpha _{j}$. Then we can proceed to study representations in terms of the
spectral function, as derived for the case of temperature by Dolan and
Jackiw~\cite{3dj}. We first consider one-compactified dimension as an
example, following the detailed calculation presented in the Appendix.

Take the topology $\Gamma _{D}^{1}$ where the imaginary time-axis is
compactified. In this case, we denote, $\alpha =(\beta ,0,\dots ,0)=\beta
\widehat{n}_{0}$, $\widehat{n}_{0}=(1,0,\dots ,0)$, with $T=\beta ^{-1}$
being the temperature, such that the Green function is given by
\begin{equation*}
G_{0}(x-y;\beta )=\frac{1}{i\beta }\sum\limits_{l_{0}}\int \frac{d^{D-1}%
\mathbf{k}}{(2\pi )^{D-1}}\frac{e^{-ik_{l_{0}}\cdot (x-y)}}{%
(k_{l_{0}})^{2}-m^{2}+i\varepsilon },
\end{equation*}%
where $k_{l_{0}}=(k_{l_{0}}^{0},k^{1},\dots ,k^{D-1})$, with $%
k_{l_{0}}^{0}=2\pi l_{0}/\beta $, being the Matsubara frequency. Then this
propagator is mapped, by an analytical continuation, into a Fourier integral
representation given by
\begin{equation}
G_{0}(x-y;\beta )=\int \frac{d^{D}k}{(2\pi )^{D}}e^{-ik(x-y)}\ G_{0}(k;\beta
),  \label{bo1012}
\end{equation}%
where%
\begin{equation*}
G_{0}(k;\beta )=G_{0}(k)+\ f_{\beta }(k^{0})[G_{0}(k)-G_{0}^{\ast }(k)]
\end{equation*}%
and%
\begin{equation*}
f_{\beta }(k^{0})=\sum\limits_{l_{0}=1}^{\infty }e^{-\beta \omega _{k}l_{0}}=%
\frac{1}{e^{\beta \omega _{k}}-1}\equiv n(k^{0};\beta ),
\end{equation*}%
which is the boson distribution function at temperature $T$ with $\omega
_{k}=k^{0}$. Then we have
\begin{equation*}
G_{0}(k;\beta )=\frac{-1}{k^{2}-m^{2}+i\varepsilon }+n(k^{0};\beta )2\pi
i\delta (k^{2}-m^{2}),
\end{equation*}

In the case of compactification of the coordinate $x^{1}$, for the topology $%
\Gamma _{D}^{1}$, we take $\alpha =(0,iL_{1},0,\dots ,0)=iL_{1}\widehat{n}%
_{1}$, with $\widehat{n}_{1}=(0,1,0,\dots ,0)$. The factor $i$ in the
parameter $\alpha _{j}$ corresponding to the compactification of a space
coordinate makes explicit that we are working with the Minkowski metric; the
period in the $x^{1}$ direction is real and equal to $L_{1}$. The propagator
has the Matsubara representation
\begin{equation*}
G_{0}(x-y;L_{1})=\frac{1}{L_{1}}\sum\limits_{l_{1}}\int \frac{d^{D-1}\mathbf{%
k}}{(2\pi )^{D-1}}\frac{e^{-ik_{l_{1}}\cdot (x-y)}}{(k_{l_{1}})^{2}-m^{2}+i%
\varepsilon },
\end{equation*}%
where $k_{l_{1}}=(k^{0},k_{l_{1}}^{1},k^{2},\dots ,k^{D-1})$, with $%
k_{l_{1}}^{1}=2\pi l_{1}/L_{1}$. The Fourier-integral representation can be
derived along the same way as in the case of temperature~\cite{kha1},
leading to
\begin{equation*}
G_{0}(x-y;L_{1})=\int \frac{d^{D}k}{(2\pi )^{D}}e^{-ik(x-y)}\ G_{0}(k;L_{1}),
\end{equation*}%
where
\begin{equation*}
G_{0}(k;L_{1})=\frac{-1}{k^{2}-m^{2}+i\varepsilon }+f_{L_{1}}(k^{1})2\pi
i\delta (k^{2}-m^{2}),
\end{equation*}%
with
\begin{equation*}
f_{L_{1}}(k^{1})=\sum\limits_{l_{1}=1}^{\infty }e^{-iL_{1}k^{1}l_{1}}.
\end{equation*}

As another example, we consider the topology $\Gamma _{D}^{2}$, accounting
for a double compactification, one being the imaginary time and the other
the $x^{1}$ direction. In this case $\alpha =(\beta ,iL_{1},0,\dots
,0)=\beta \widehat{n}_{0}+iL_{1}\widehat{n}_{1}$. The Matsubara
representation is
\begin{equation*}
G_{0}(x-y;\beta ,L_{1})=\frac{1}{i\beta L_{1}}\sum\limits_{l_{0},l_{1}}\int
\frac{d^{D-2}\mathbf{k}}{(2\pi )^{D-2}}\frac{e^{-ik_{l_{0}l_{1}}\cdot
(x-x^{\prime })}}{(k_{l_{0}l_{1}})^{2}-m^{2}+i\varepsilon },
\end{equation*}%
where $k_{l_{0}l_{1}}=(k_{l_{0}}^{0},k_{l_{1}}^{1},k^{2},\dots
,k^{D-1})$, with $k_{l_{0}}^{0}=2\pi l_{0}/\beta $ and
$k_{l_{1}}^{1}=2\pi l_{1}/L_{1}$. The corresponding Fourier-integral
representations is
\begin{equation*}
G_{0}(x-y;\beta ,L_{1})=\int \frac{d^{D}k}{(2\pi )^{D}}e^{-ik(x-y)}\
G_{0}(k;\beta ,L_{1}),
\end{equation*}%
where
\begin{equation*}
G_{0}(k;\beta ,L_{1})=\frac{-1}{k^{2}-m^{2}+i\varepsilon }+f_{\beta
L_{1}}(k^{0},k^{1})2\pi i\delta (k^{2}-m^{2}),
\end{equation*}%
with
\begin{equation*}
f_{\beta L_{1}}(k^{0},k^{1})=f_{\beta }(k^{0})+f_{L_{1}}(k^{1})+2f_{\beta
}(k^{0})f_{L_{1}}(k^{1}).
\end{equation*}%
In the Appendix we demonstrate this result, as well as the generalization
for $d$ compactified dimensions. In any case, the general structure of the
propagator is given by
\begin{equation}
G_{0}(x-y;\alpha )=\int \frac{d^{D}k}{(2\pi )^{D}}e^{-ik(x-y)}\
G_{0}(k;\alpha ),  \label{bo10121}
\end{equation}%
where%
\begin{equation}
G_{0}(k;\alpha )=G_{0}(k)+\ f_{\alpha }(k_{\alpha })[G_{0}(k)-G_{0}^{\ast
}(k)],  \label{CBPF22.12.3}
\end{equation}

This is one of the main result in this paper. The next step is to consider
interacting systems in a topology $\Gamma _{D}^{d}$. Before that, we have to
finish the GNS construction for this case.

\section{$^{\ast }$-Lie algebras and field theory}

Our main goal in this section is to demonstrate the existence of the states $%
|\xi _{w}^{\alpha }\rangle $, introduced in Eq.~(\ref{CBPF1}), as a second
part of the GNS construction. For that, we study first general elements of
representations for Lie algebras, by using the modular representations of a
c*-algebra. Applying the results for the Poincar\'{e} group, we analyze
representations describing free bosons and fermions compactified in
space-time.

\subsection{Modular representation and Lie groups}

Consider $\ell =\{a_{i},i=1,2,3,...\}$ a Lie algebra over the (real) field $%
\mathbb{R}$, of a Lie group $\mathcal{G}$, characterized by the algebraic
relations $(a_{i},a_{j})=C_{ijk}a_{k}$, where $C_{ijk}\in \mathbb{R}$ are
the structure constants and $(,)$ is the Lie product (we are assuming the
convention of summation over repeated indices). Using the modular
conjugation, $^{\ast }$-representations for $\ell ,$ denoted by $^{\ast
}\ell $, are constructed. Let us take $\pi (\ell ),$ a representation for $%
\ell $ as a von Neumann algebra, and $\widetilde{\pi }(\ell )$ as the
representation for the correspondent commutant. Each element in $\ell $ is
denoted by $\pi (a_{i})=A_{i}$ and $\widetilde{\pi }(a_{i})=\widetilde{A}
_{i} $; thus we have~\cite{ade1},
\begin{align}
\lbrack {\tilde{A}}_{i},{\tilde{A}}_{j}]& =-iC_{ijk}{\tilde{A}}_{k},
\label{oct1} \\
\lbrack A_{i},A_{j}]& =iC_{ijk}A_{k},  \label{oct2} \\
\lbrack {\tilde{A}}_{i},A_{j}]& =0.  \label{oct3}
\end{align}%
This provides a reducible representation for $\ell $ without an apparent
physical or mathematical outcome of interest. However, a careful analysis
brings out facts that are important, at least, in physics. The modular
generators of symmetry are given by $\widehat{A}=A-\widetilde{A}.$ Then we
have from Eqs.~(\ref{oct1})-(\ref{oct3}) that the $^{\ast }\ell $ algebra is
given by%
\begin{align}
\lbrack \widehat{A}_{i},\widehat{A}_{j}]& =iC_{ijk}\widehat{A}_{k},
\label{aug09141} \\
\lbrack \widehat{A}_{i},A_{j}]& =iC_{ijk}A_{k},  \label{aug09142} \\
\lbrack A_{i},A_{j}]& =iC_{ijk}A_{k}.  \label{aug09143}
\end{align}%
This is just the semidirect product of the faithful representation $\pi
(a_{i})=A_{i}$ and the other faithful representation $\widehat{\pi }(A_{i})=
\widehat{A}_{i},$ with $\pi (a_{i})$ providing elements of the invariant
subalgebra. This is the proof of the following statement:

\begin{itemize}
\item Proposition 2. Consider the Tomita-Takesaki representation, where the
von Neumann algebra is a Lie algebra, $\ell $. Then the modular
representation for $\ell $ is given by Eqs.~(\ref{aug09141})-(\ref{aug09143}
), the $^{\ast }\ell $-algebra, where the invariant subalgebra describes
properties of observables of the theory, that are transformed under the
symmetry defined by the generators of modular transformations.
\end{itemize}

Another aspect to be explored is a set of linear mappings $U(\xi ):\pi _{w}(%
\mathcal{A})\times \widetilde{\pi }_{w}(\mathcal{A})\rightarrow \pi _{w}(%
\mathcal{A})\times \widetilde{\pi }_{w}(\mathcal{A})$ with the
characteristics of a Bogoliubov transformation, i.e. $U(\xi )$ is canonical,
in the sense of keeping the algebraic relations, and unitary but only for a
finite dimensional basis. Then we have a group with elements $U(\xi )$
specified by the parameters\ $\xi $. \ This is due to the two commutant sets
in the von Neumann algebra. The characteristic of $U(\xi )$ as a linear
mapping is guaranteed by the canonical invariance of $^{\ast }\ell .$ In
terms of generators of symmetry and tilde operators we obtain,
\begin{align*}
A(\xi )& =U(\xi )AU(\xi )^{-1}, \\
\widetilde{A}(\xi )& =U(\xi )\widetilde{A}U(\xi )^{-1},
\end{align*}%
such that%
\begin{align*}
\lbrack {\tilde{A}(\xi )}_{i},{\tilde{A}(\xi )}_{j}]& =-iC_{ijk}{\tilde{A}%
(\xi )}_{k}, \\
\lbrack A(\xi )_{i},A(\xi )_{j}]& =iC_{ijk}A(\xi )_{k}, \\
\lbrack {\tilde{A}(\xi )}_{i},A{(\xi )}_{j}]& =0.
\end{align*}

The goal here is to use $U(\xi )$ to construct explicitly the states $%
w_{A}^{\xi }(x)$ introduced in Eq.~(\ref{adol13}), describing fields in a $%
\Gamma _{D}^{d}$ topology. For the Poincar\'{e} algebra, for instance, we
have the $^{\ast }$-Poincar\'{e} Lie algebra ($^{\ast }\mathfrak{p}$) given
by
\begin{align}
\lbrack M_{\mu \nu },P_{\sigma }]& =i(g_{\nu \sigma }P_{\mu }-g_{\sigma \mu
}P_{\nu }),  \label{poin1} \\
\lbrack P_{\mu },P_{\nu }]& =0,  \label{poin2} \\
\lbrack M_{\mu \nu },M_{\sigma \rho }]& =-i(g_{\mu \rho }M_{\nu \sigma
}-g_{\nu \rho }M_{\mu \sigma }+g_{\mu \sigma }M_{\rho \nu }-g_{\nu \sigma
}M_{\rho \mu }),  \label{poin3} \\
\lbrack \widetilde{M}_{\mu \nu },\widetilde{P}_{\sigma }]& =-i(g_{\nu \sigma
}\widetilde{P}_{\mu }-g_{\sigma \mu }\widetilde{P}_{\nu }),  \label{poin7} \\
\lbrack \widetilde{P}_{\mu },\widetilde{P}_{\nu }]& =0,  \label{poin8} \\
\lbrack \widetilde{M}_{\mu \nu },\widetilde{M}_{\sigma \rho }]& =i(g_{\mu
\rho }\widetilde{M}_{\nu \sigma }-g_{\nu \rho }\widetilde{M}_{\mu \sigma
}+g_{\mu \sigma }\widetilde{M}_{\rho \nu }-g_{\nu \sigma }\widetilde{M}%
_{\rho \mu }),  \label{poin9}
\end{align}%
where $\widetilde{M}_{\mu \nu }=\widetilde{M}_{\mu \nu }(\xi )$, $M_{\mu \nu
}=M_{\mu \nu }(\xi ),$ $\widetilde{P}_{\mu }=\widetilde{P}_{\mu }(\xi )$ and
$P_{\mu }=P_{\mu }(\xi )$. All other commutation relations are zero. For this
algebra, we obtain representations for generators of symmetry. Generators of
the Poincar\'{e} symmetry are given by $\widehat{M}_{\mu \nu }=M_{\mu \nu }-%
\widetilde{M}_{\mu \nu }$ and $\widehat{P}_{\mu }=P_{\mu }-\widetilde{P}%
_{\mu }$ and satisfy the commutation relations similar to those given by
Eqs.~(\ref{aug09141})-(\ref{aug09143}). The representations are constructed
by using a set of Casimir invariants, i.e.
\begin{align}
w^{2}& =w_{\mu }w^{\mu },  \label{inv1} \\
P^{2}& =P_{\mu }P^{\mu },  \label{inv2} \\
\widetilde{w^{2}}& =\widetilde{w}_{\mu }\widetilde{w}^{\mu },  \label{inv3}
\\
\widetilde{P^{2}}& =\widetilde{P}_{\mu }\widetilde{P}^{\mu }  \label{inv4}
\end{align}%
where $w_{\mu }=\frac{1}{2}\varepsilon _{\mu \nu \sigma \rho }M^{\nu \sigma
}P^{\rho }$ is the Pauli-Lubanski vector.

\subsection{Boson fields}

Let us consider a free quantum field describing bosons. The modular
conjugation rules can be applied to any relation among the dynamical
variables, in particular to the equation of motion in the Heisenberg
picture. The set of doubled equations are then derived by writing the
hat-Hamiltonian, the generator of time translation, as $\widehat{H}=H-%
\widetilde{H}$. However, due to the GNS construction, we need to consider
only the evolution of the Lagrangian for non-tilde operators, evolving in
space-time by the generators of $^{\ast }\mathfrak{p}$. In this case the
time evolution generator is $\widehat{H}.$ Then we have the Lagrangian
densities $\mathcal{L}(x)$ and $\mathcal{L}(x;\xi )$ given, respectively, by
\begin{align}
\mathcal{L}(x)& =\frac{1}{2}\partial _{\mu }\phi (x)\partial ^{\mu }\phi (x)-%
\frac{m^{2}}{2}\phi (x)^{2},  \label{may10202} \\
\mathcal{L}(x;\xi )& =\frac{1}{2}\partial _{\mu }\phi (x;\xi )\partial ^{\mu
}\phi (x;\xi )-\frac{m^{2}}{2}\phi (x;\xi )^{2},  \label{aug091441}
\end{align}%
where the field $\phi (x;\xi )$ is defined by
\begin{equation*}
\phi (x;\xi )=U(\xi )\phi (x)U^{-1}(\xi ).
\end{equation*}

The mapping $U(\xi )$ is taken as a Bogoliubov transformation and is
defined, as usual, by a two-mode squeezed operator. For fields expanded in
terms of modes, we define
\begin{align}
U(\xi )& =\exp \left\{ \sum_{k}\theta (k_{\xi };\xi )[a^{\dag }(k)\widetilde{%
a}^{\dag }(k)-a(k)\widetilde{a}(k)]\right\}  \notag \\
& =\prod_{k}U(k;\xi ),  \label{bog731}
\end{align}%
where
\begin{equation*}
U(k_{\xi };\xi )=\exp \{\theta (k_{\xi };\xi )[a^{\dag }(k)\widetilde{a}%
^{\dag }(k)-a(k)\widetilde{a}(k)]\},
\end{equation*}%
with $\theta (k_{\xi };\xi )$ being a function of the momentum, $k_{\xi }$,
and of the parameters $\xi ,$ both to be specified. The label $k$ in the sum
and in the product of the equations above is to be taken in the continuum
limit, for each mode. Then we have
\begin{equation}
\phi (x;\xi )=\int \frac{d^{D-1}\mathbf{k}}{(2\pi )^{D-1}}\frac{1}{2k_{0}}%
[a(k;\xi )e^{-ikx}+a^{\dag }(k;\xi )e^{ikx}]. \label{xifield}
\end{equation}%
To obtain this expression, we have used the non-zero commutation relations
\begin{equation}
\lbrack a(k;\xi ),a^{\dag }(k^{\prime };\xi )]=(2\pi )^{3}2k_{0}\delta (%
\mathbf{k}-\mathbf{k}^{\prime }),  \label{com73}
\end{equation}%
with
\begin{align*}
a(k;\xi )& =U(k_{\xi };\xi )a(k)U^{-1}(k_{\xi };\xi ) \\
& =u(k_{\xi };\xi )a(k)-v(k_{\xi };\xi )\,\widetilde{a}^{\dagger }(k),
\end{align*}%
where $u(k_{\xi };\xi )$ and $v(k_{\xi };\xi )$ are given in terms of $%
\theta (k_{\xi };\xi )$ by
\begin{align*}
u(k_{\xi };\xi )& =\cosh \theta (k_{\xi };\xi ), \\
v(k_{\xi };\xi )& =\sinh \theta (k_{\xi };\xi ).
\end{align*}%
The inverse is
\begin{equation*}
a(k)=u(k_{\xi };\xi )a(k;\xi )+v(k_{\xi };\xi )\,\widetilde{a}^{\dagger
}(k;\xi ),
\end{equation*}%
such that the other operators $a^{\dag }(k),\widetilde{a}(k)$ and $%
\widetilde{a}^{\dagger }(k)$ are obtained by applying the hermitian
conjugation or the tilde conjugation, or both.

It is worth noting that the transformation $U(\xi )$ can be mapped into a $%
2\times 2$ representation of the Bogoliubov transformation, i.e.%
\begin{equation}
B(k_{\xi };\xi )=\left(
\begin{array}{cc}
u(k_{\xi };\xi ) & -v(k_{\xi };\xi ) \\
-v(k_{\xi };\xi ) & u(k_{\xi };\xi )%
\end{array}%
\right) ,  \label{Bmar101}
\end{equation}%
with $u^{2}(k_{\xi };\xi )-v^{2}(k_{\xi };\xi )=1$, acting on the pair of
commutant operators as
\begin{equation*}
\left(
\begin{array}{c}
a(k;\xi ) \\
\widetilde{a}^{\dagger }(k;\xi )%
\end{array}%
\right) =B(k_{\xi };\xi )\left(
\begin{array}{c}
a(k) \\
\widetilde{a}^{\dagger }(k)%
\end{array}%
\right) .
\end{equation*}%
A Bogoliubov transformation of this type gives rise to a compact and elegant
$2\times 2$ representation of the propagator in the real-time formalism.
However, to derive and use a quantum field theory in a topology $\Gamma
_{D}^{d}$ following the GNS construction, we observe that this matrix
representation for the propagator is indeed not necessary. This aspect is
useful for applications, in particular to represent an ease in the
calculations of physical processes.

The Hilbert space is constructed from the $\xi $-state, $|0(\xi )\rangle
=U(\xi )|0,\widetilde{0}\rangle ,$ where $|0,\widetilde{0}\rangle =\underset{%
k}{\bigotimes }|0,\widetilde{0}\rangle _{k}$ and $|0,\widetilde{0}\rangle
_{k}$ is the vacuum for the mode $k.$ Then we have: $a(k;\xi )|0(\xi
)\rangle =\widetilde{a}(k;\xi )|0(\xi )\rangle =0$ and $\langle 0(\xi
)|0(\xi )\rangle =1.$ This shows that $|0(\xi )\rangle $ is a vacuum for $%
\xi $-operators $a(k;\xi )$. However, it is a condensate for the operators $%
a $ and $a^{\dagger }.$ An arbitrary basis vector is given in the form
\begin{eqnarray}
|\psi (\xi );\{m\};\{k\}\rangle &=&[a^{\dag }(k_{1};\xi )]^{m_{1}}\cdots
\lbrack a^{\dag }(k_{M};\xi )]^{m_{M}}  \notag \\
&&\times \lbrack \widetilde{a}^{\dag }(k_{1};\xi )]^{n_{1}}\cdots \lbrack
\widetilde{a}^{\dag }(k_{N};\xi )]^{n_{N}}|0(\xi )\rangle ,
\label{therm12345}
\end{eqnarray}%
where $n_{i},m_{j}=0,1,2,...,$ with $N$ and $M$ being indices for an
arbitrary mode.

Consider only one field-mode, for simplicity. Then we write $|0(\xi )\rangle
$ in terms of $u(\xi )$ and $v(\xi )$ as
\begin{align}
|0(\xi )\rangle & =\frac{1}{u(\xi )}\exp [\frac{v(\xi )}{u(\xi )}a^{\dagger }%
\widetilde{a}^{\dagger }]|0,\widetilde{0}\rangle   \notag \\
& =\frac{1}{u(\xi )}\sum\limits_{n}\left( \frac{v(\xi )}{u(\xi )}\right)
^{n}|n,\widetilde{n}\rangle .  \label{vacuu12}
\end{align}%
This provides an explicit example of states in the GNS construction for a
quantum field theory in a topology $\Gamma _{D}^{d}$. Since the state $%
|0(\xi )\rangle $ is a trace-like state, this leads to the state $%
w_{A}^{\alpha }(x)$. At this point, the physical meaning of an arbitrary $%
\xi $-state given in Eq.~(\ref{therm12345}) is not established. This aspect
is discussed by considering the Green function defined by
\begin{equation*}
G_{0}(x-y;\xi )=-i\langle 0(\xi )|\mathrm{T}[\phi (x)\phi (y)]|0(\xi
)\rangle .
\end{equation*}%
We demand then that $G_{0}(x-y;\xi )\equiv G_{0}(x-y;\alpha )$, where $%
G_{0}(x-y;\alpha )$ is given in Eq. (\ref{aug09148}). Using $U(\xi )$ in
Eq.~(\ref{bog731}), we find that the $\xi $-Green function is written as
\begin{equation*}
G_{0}(x-y;\xi )=-i\langle \widetilde{0},0|\mathrm{T}[\phi (x;\xi )\phi
(y;\xi )]|0,\widetilde{0}\rangle .
\end{equation*}%
Then, using the field expansion (\ref{xifield}) and the commutation relation (\ref{com73}),
we obtain
\begin{equation}
G_{0}(x-y;\xi )=\int \frac{d^{D}k}{(2\pi )^{D}}e^{-ik(x-y)}\ G_{0}(k;\xi ),
\label{bo10}
\end{equation}%
where%
\begin{equation*}
G_{0}(k;\xi )=G_{0}(k)+\ v^{2}(k_{\xi };\xi )[G_{0}(k)-G_{0}^{\ast }(k)].
\end{equation*}%

This propagator is formally identical to $G_{0}(x-y;\alpha )$
written in the integral representation given by
Eqs.~(\ref{bo10121}) and (\ref{CBPF22.12.3}). Then the analysis in terms of
representation of Lie-groups and the Bogoliubov transformation
leads to the integral representation by performing the mapping
$v^{2}(k_{\xi };\xi )\rightarrow f_{\alpha }(k_{\alpha })$. Notice
that this is possible, since $v^{2}(k_{\xi };\xi )$ has not been
fully specified up to this point. Considering the specific case of
compactification in time, in order to describe temperature only,
the real quantity $v^{2}(k_{\xi };\xi )$ is mapped in the real
quantity $ f_{\beta }(w )\equiv n(\beta)$. Including space
compactification, $ f_{\alpha }(k_{\alpha })$ is a complex
function, according to Appendix. In such a case, we can consider $
f_{\alpha }(k_{\alpha })$ as the analytical continuation of the
 real function $v^{2}(k_{\xi };\xi )$; a procedure that is
 possible, since, $v^{2}(k_{\xi };\xi )$ is arbitrary. Therefore,
for space compactification, we can also perform the mapping
$v^{2}(k_{\xi };\xi )\rightarrow f_{\alpha }(k_{\alpha })$ in
$G_{0}(k;\xi )$, in order to recover the propagator shown in
Eqs.~(\ref{bo10121}) and (\ref{CBPF22.12.3}). From now on, we denote the vector
$|\xi^{\alpha}_{w}\rangle$  by $|\alpha\rangle$ and the function $f_{\alpha }(k_{\alpha })$
by $v^{2}(k_{\alpha };\alpha)$.

\subsection{Fermion field}

A similar mathematical structure is introduced for the compactification of
fermion fields. With the average given in Eq.~(\ref{adol11}), $w_{A}^{\alpha
}(x)\equiv \langle \alpha |A(x)|\alpha \rangle $, $A(x)$ is defined in terms
of fermion operators. We have first to construct the state $|\alpha \rangle $
explicitly.

The Lagrangian density for the free Dirac field is
\begin{equation}
\mathcal{L}(x)=\frac{1}{2}\overline{\psi }(x)\left[\gamma \cdot i
\overleftrightarrow{\partial }-m \right]\psi (x)
\end{equation}
and for the $\alpha $-field we have
\begin{equation}
\mathcal{L}(x;\alpha )=\frac{1}{2}\overline{\psi }(x;\alpha ) \left[\gamma
\cdot i \overleftrightarrow{\partial }-m \right]\psi (x;\alpha ).
\end{equation}
The field $\psi (x;\alpha )$ is expanded as
\begin{equation*}
\psi (x;\alpha )=\int \frac{d^{D-1}k}{(2\pi )^{D-1}}\frac{m}{k_{0}}
\sum\limits_{\zeta =1}^{2}\left[c_{\zeta }(k;\alpha )u^{(\zeta
)}(k)e^{-ikx}+d_{\zeta }^{\dagger }(k;\alpha )v^{(\zeta )}(k)e^{ikx}\right],
\end{equation*}
where $u^{(\zeta )}(k)$ and $v^{(\zeta )}(k)$ are basic spinors. The fermion
field $\psi (x;\alpha )$ is defined by
\begin{equation*}
\psi (x;\alpha )=U(\alpha )\psi (x)U^{-1}(\alpha ),
\end{equation*}%
where $U(\alpha )$ is
\begin{eqnarray*}
U(\alpha ) &=&\exp \left\{ \sum_{k}\{\theta _{c}(k;\alpha )[c^{\dag }(k)
\widetilde{c}^{\dag }(k)-c(k)\widetilde{c}(k)] \right. \\
& & \left. \;\;\;\;\;\;\;\;\;\;\;\;\;\;+\theta _{d}(k;\alpha )[d^{\dag }(k)%
\widetilde{d}^{\dag }(k)-d(k)\widetilde{d}(k)]\}\right\} \\
&=&\prod\limits_{k}U_{c}(k;\alpha )U_{d}(k;\alpha ),
\end{eqnarray*}%
with
\begin{eqnarray*}
U_{c}(k;\alpha ) &=&\exp \{\theta _{c}(k;\alpha )[c^{\dag }(k)\widetilde{c}
^{\dag }(k)-c(k)\widetilde{c}(k)]\}, \\
U_{d}(k;\alpha ) &=&\exp \{\theta _{d}(k;\alpha )[d^{\dag }(k)\widetilde{d}
^{\dag }(k)-d(k)\widetilde{d}(k)]\}.
\end{eqnarray*}%
The fermion $\alpha $-operators $c(k;\alpha )$ and $d(k;\alpha )$ are
written in terms of non $\alpha $-operators by
\begin{eqnarray*}
c(k;\alpha ) &=&U(\alpha )c(k)U^{-1}(\alpha )=U(k;\alpha
)c(k)U^{-1}(k;\alpha ) \\
&=&u_{c}(k;\alpha )c(k)-v_{c}(k;\alpha )\widetilde{c}^{\dagger }(k), \\
d(k;\alpha ) &=&U(\alpha )d(k)U^{-1}(\alpha )=U(k;\alpha
)d(k)U^{-1}(k;\alpha ) \\
&=&u_{d}(k;\alpha )d(k)-v_{d}(k;\alpha )\widetilde{d}^{\dagger }(k).
\end{eqnarray*}
The parameters $\theta _{c}(k;\alpha )$ and $\theta _{d}(k;\alpha )$ are
such that $\sin \theta _{c}(k;\alpha )=v_{c}(k_{\alpha };\alpha )$, and $%
\sin \theta _{d}(k;\alpha )=v_{d}(k_{\alpha };\alpha )$, resulting in $%
v_{c}^{2}(k_{\alpha };\alpha )+u_{c}^{2}(k_{\alpha };\alpha )=1$ and $%
v_{d}^{2}(k_{\alpha };\alpha )+u_{d}^{2}(k_{\alpha };\alpha )=1$. The
inverse formulas for the $\alpha $-operators are
\begin{eqnarray*}
c(k) &=&u_{c}(k_{\alpha };\alpha )c(k;\alpha )+v_{c}(k_{\alpha };\alpha )
\widetilde{c}^{\dagger }(k;\alpha ), \\
d(k) &=&u_{d}(k_{\alpha };\alpha )d(k;\alpha )+v_{d}(k;\alpha )\widetilde{d}
^{\dagger }(k;\alpha ).
\end{eqnarray*}
Observe that the operators $c$ and $d$ carry a spin index.

These operators satisfy the anti-commutation relations
\begin{equation*}
\{c_{\zeta }(k,\alpha \mathbf{),}c_{\varkappa }^{\dagger }(k^{\prime
},\alpha \mathbf{)}\}=\{d_{\zeta }(k,\alpha \mathbf{),}d_{\varkappa
}^{\dagger }(k^{\prime },\alpha \mathbf{)}\}=(2\pi )^{3}\frac{k_{0}}{m}
\delta (\mathbf{k-k}^{\prime })\delta _{\zeta \varkappa },
\end{equation*}%
with all the other anti-commutation relations being zero. In order to be
consistent with the Lie algebra, and with the definition of the $\alpha $
-operators, a fermion operator, $A,$ is such that $\widetilde{\widetilde{A}}
=-A$ and a tilde-fermion operator anti-commutes with a non-tilde operator.
This is consistent in the following sense. Consider, for instance, $%
c(k;\alpha )=U(k;\alpha )c(k)U^{-1}(k;\alpha )$ and $\widetilde{c}(k;\alpha
)=U(k;\alpha )\widetilde{c}(k)U^{-1}(k;\alpha )$. In order to map $%
c(k;\alpha )\rightarrow \widetilde{c}(k;\alpha )$ by using the modular
conjugation, directly, it leads to $\widetilde{\widetilde{c}}(k)=-c(k)$.
This is important to preserve the canonical structure of $U(k;\alpha )$,
regarding in particular the $^{\ast }$Lie-algebra. This analysis provides
then a precise and simple way to define the modular conjugation for fermions.

Let us define the $\alpha $-state $|0(\alpha )\rangle =U(\alpha )|0,\tilde{0}
\rangle ,$ where
\begin{equation*}
|0,\tilde{0}\rangle =\bigotimes_{k}|0,\tilde{0}\rangle _{k}
\end{equation*}%
and $|0,\tilde{0}\rangle _{k}$ is the vacuum for the mode $k$ for particles
and anti-particles. This $\alpha $-state satisfies the condition $\langle
0(\alpha )|0(\alpha )\rangle =1$. Moreover, we have%
\begin{eqnarray*}
c(k;\alpha )|0(\alpha )\rangle &=&\widetilde{c}(k;\alpha )|0(\alpha )\rangle
=0, \\
d(k;\alpha )|0(\alpha )\rangle &=&\widetilde{d}(k;\alpha )|0(\alpha )\rangle
=0.
\end{eqnarray*}%
Then $|0(\alpha )\rangle $ is a vacuum state for the $\alpha $-operators $%
c(k;\alpha )$ and $d(k;\alpha )$. Basis vectors are given in the form
\begin{equation*}
\lbrack c^{\dag }(k_{1};\alpha )]^{r_{1}}\cdots [d^{\dag }(k_{M};\alpha
)]^{r_{M}}[\widetilde{c}^{\dag }(k_{1};\alpha )]^{s_{1}}\cdots [\widetilde{d}
^{\dag }(k_{N};\alpha )]^{s_{N}}|0(\alpha )\rangle ,
\end{equation*}%
where $r_{i},s_{i}=0,1$. A general $\alpha $-state can then be defined by a
linear combinations of such basis vectors.

Let us consider some particular cases, first, the case of temperature. The
topology is $\Gamma _{D}^{1}$, and we take $\alpha =(\beta ,0,\dots,0)$,
leading to
\begin{eqnarray*}
v_{c}^{2}(k^{0};\beta ) &=&\frac{1}{e^{\beta (w_{k}-\mu _{c})}+1}, \\
v_{d}^{2}(k^{0};\beta ) &=&\frac{1}{e^{\beta (w_{k}+\mu _{d})}+1},
\end{eqnarray*}%
where $\mu _{c}$ and $\mu _{d}$ are the chemical potential for particles and
antiparticles, respectively. For simplicity, we take $\mu _{c}=\mu _{d}=0$, and write $%
v_{F}(k^{0};\beta )=v_{c}(k^{0};\beta )=v_{d}(k^{0};\beta )$, such that
\begin{equation*}
v_{F}^{2}(k^{0};\beta )=\frac{1}{e^{\beta w_{k}}+1}=\sum\limits_{n=1}^{
\infty }(-1)^{1+n}e^{-\beta w_{k}n}.
\end{equation*}%
For the case of spatial compactification, we take $\alpha
=(0,iL_{1},0,\dots,0)$. By a kind of Wick rotation, we derive $%
v_{F}^{2}(k^{1};L_{1}) $ from $v_{F}^{2}(k^{0};\beta )$, resulting in%
\begin{equation*}
v_{F}^{2}(k^{1};L_{1})=\sum\limits_{n=1}^{\infty }(-1)^{1+n}e^{-iL_{1}k^{1}n}.
\end{equation*}%
For spatial compactification and temperature, we have (see the Appendix)
\begin{equation*}
v_{F}^{2}(k^{0},k^{1};\beta ,L_{1})=v_{F}^{2}(k^{1};\beta
)+v_{F}^{2}(k^{1};L_{1})+2v_{F}^{2}(k^{1};\beta )v_{F}^{2}(k^{1};L_{1}).
\end{equation*}

The $\alpha $-Green function is defined by $S_{0}(x,y;\alpha )=w_{A}^{\alpha
}(x,y)\equiv \langle 0(\alpha )|A(x,y)|0(\alpha )\rangle $, where $A(x,y)=
\mathrm{T}[\psi (x)\overline{\psi }(y)]$. Then we have
\begin{equation}
S_{0}(x-y;\alpha )=-i\langle 0(\alpha )|\mathrm{T}[\psi (x)\overline{\psi }
(y)]|0(\alpha )\rangle .  \label{prop78}
\end{equation}%
Let us write%
\begin{equation}
iS_{0}(x-y;\alpha )=\theta (x^{0}-y^{0})S(x-y;\alpha )-\theta (y^{0}-x^{0})
\overline{S}(y-x;\alpha ),  \label{feyn79}
\end{equation}%
with $S(x-y;\alpha )=\langle 0(\alpha )|\psi (x)\overline{\psi }(y)|0(\alpha
)\rangle $ and $\overline{S}(x-y;\alpha )=\langle 0(\alpha )|\overline{\psi }
(y)\psi (x)|0(\alpha )\rangle $. Calculating $S$ and $\overline{S}$ we
obtain
\begin{eqnarray*}
S(x-y;\alpha ) &=&(i\gamma \cdot \partial +m)\int \frac{d^{D-1}k}{(2\pi
)^{D-1}}\frac{1}{2\omega_k}  \notag \\
&&\times \lbrack e^{-ik(x-y)}- v_{F}^{2}(k_{\alpha };\alpha
)(e^{-ik(x-y)}-e^{ik(x-y)})],  \label{prop713}
\end{eqnarray*}%
For the term $\overline{S}(x-y;\alpha )$, we have
\begin{eqnarray*}
\overline{S}(x-y;\alpha ) &=&(i\gamma \cdot \partial +m)\int \frac{d^{D-1}k}{
(2\pi )^{D-1}} \frac{1}{2\omega_k}  \notag \\
&&\times \lbrack - e^{ik(x-y)}+ v_{F}^{2}(k_{\alpha };\alpha
)(e^{-ik(x-y)}+e^{ik(x-y)})].  \label{prop724}
\end{eqnarray*}%
This leads to%
\begin{equation}
S_{0}(x-y;\alpha )=(i\gamma \cdot \partial +m)G_{0}^{F}(x-y;\alpha ), \label{Sfermions}
\end{equation}%
where%
\begin{equation}
G_{0}^{F}(x-y;\alpha )=\int \frac{d^{D}k}{(2\pi )^{D}}e^{-ik(x-y)}
G_{0}^{F}(k;\alpha ), \label{Gxfermi}
\end{equation}%
and%
\begin{equation}
G_{0}^{F}(k;\alpha )=G_{0}(k)+\ v_{F}^{2}(k_{\alpha };\alpha
)[G_{0}(k)-G_{0}^{\ast }(k)]. \label{Gkfermi}
\end{equation}%

This Green function is similar to the boson Green function, Eq.~(\ref{bo10}%
); the difference is the fermion function $v_{F}^{2}(k_{\alpha };\alpha )$.
Again we observe that, due to the GNS construction, the $2\times 2$%
-representation of the propagator is not necessary, although it can be
introduced. In such a case the Bogoliubov transformation is written in the
form of a $2\times 2$ matrix for particles (subindex $c$) and anti-particles
(subindex $d$) is
\begin{equation}
B_{c,d}(k_{\alpha };\alpha )=\left(
\begin{array}{cc}
u_{c,d}(k_{\alpha };\alpha ) & v_{c,d}(k_{\alpha };\alpha ) \\
-v_{c,d}(k_{\alpha };\alpha ) & u_{c,d}(k_{\alpha };\alpha )%
\end{array}%
\right) .  \label{bog23july}
\end{equation}

\section{Generating functional}

We now construct the generating functional for interacting fields living in a flat space with
topology $\Gamma _{D}^{d}$.

\subsection{Bosons}

For a system of free bosons, we consider,\ up to normalization factors, the
following generating functional
\begin{equation}
Z_{0}\simeq \int D\phi e^{iS}=\int D\phi \exp [i\int dx\mathcal{L}]=\int
D\phi \exp \{-i\int dx[\frac{1}{2}\phi (\square +m^{2})\phi -J\phi ]\},
\end{equation}
where $J$ is a source. Such a functional is written as
\begin{equation}
Z_{0}\simeq \exp \{\frac{i}{2}\int dxdy[J(x)(\square +m^{2}-i\varepsilon
)^{-1}J(y)]\},  \label{andr1}
\end{equation}%
describing the usual generating functional for bosons. However, we would
like to introduce the topology $\Gamma _{D}^{d}$. This is possible by
finding a solution of the Klein-Gordon equation
\begin{equation}
(\square +m^{2}+i\varepsilon )G_{0}(x-y;\alpha )=-\delta (x-y).
\label{KG11mar1}
\end{equation}%
Using this result in Eq.~(\ref{andr1}), we find the normalized functional%
\begin{equation}
Z_{0}[J;\alpha ]=\exp \{\frac{i}{2}\int dxdy[J(x)G_{0}(x-y;\alpha )J(y)]\}.
\label{andr4}
\end{equation}%
Then\ we have%
\begin{equation*}
G_{0}(x-y;\alpha )=i\frac{\delta ^{2}Z[J;\alpha ]}{\delta J(y)\delta J(x)}%
|_{J=0}.
\end{equation*}%
In order to treat interactions, we analyze the usual approach with the $%
\alpha $-Green function. The Lagrangian density is
\begin{equation*}
\mathcal{L}(x)=\frac{1}{2}\partial _{\mu }\phi (x)\partial ^{\mu }\phi (x)-%
\frac{m^{2}}{2}\phi ^{2}+{\mathcal{L}}_{int},
\end{equation*}%
where ${\mathcal{L}}_{int}={\mathcal{L}}_{int}(\phi )$ is the interaction
Lagrangian density. The functional $Z[J;\alpha ]$ satisfies the equation
\begin{equation*}
(\square +m)\frac{\delta Z[J;\alpha ]}{i\delta J(x)}+L_{int}\left( \frac{1}{%
i }\frac{\delta }{\delta J}\right) Z[J;\alpha ]=J(x)Z[J;\alpha ]
\end{equation*}%
with the normalized solution given by%
\begin{equation*}
Z[J;\alpha ]=\frac{\exp \left[ i\int dxL_{int}\left( \frac{1}{i}\frac{\delta
}{\delta J}\right) \right] Z_{0}[J;\alpha ]}{\exp \left[ i\int
dxL_{int}\left( \frac{1}{i}\frac{\delta }{\delta J}\right) \right]
Z_{0}[J;\alpha ]|_{J=0}}.
\end{equation*}%
Observe that the topology does not change the interaction. This is a
consequence of the isomorphism and the fact that we are considering a local
interaction. Now we turn our attention to construct the $\alpha $-generator
functional for fermions.

\subsection{Fermions}

The Lagrangian density for a free fermion system with sources is
\begin{equation*}
\mathcal{L}=i\overline{\psi }\gamma ^{\mu }\partial _{\mu }\psi -m\overline{
\psi }\psi +\overline{\psi }\eta +\overline{\eta }\psi .
\end{equation*}%
The functional $\ Z_{0}\simeq \int D\psi D\overline{\psi }e^{iS}$ is then
reduced to
\begin{equation}
\ Z_{0}[\eta ,\overline{\eta };\alpha ]=\exp \{-i\int dxdy[\overline{\eta }
(x)S_{0}(x-y;\alpha )\eta (x)]\},  \label{KG11mar2}
\end{equation}%
where%
\begin{equation*}
S_{0}(x-y;\alpha )^{-1}=\ i\gamma ^{\mu }\partial _{\mu }-m.
\end{equation*}%
Since $S_{0}(x-y;\alpha )^{-1}S_{0}(x-y;\alpha )=\delta (x-y)$, and $%
G_{0}(x-y;\alpha )$ satisfies Eq.~(\ref{KG11mar1}), we find
\begin{equation*}
S_{0}(x-y;\alpha )=(i\gamma \cdot \partial +m)G_{0}^{F}(x-y;\alpha ).
\end{equation*}%
The functional given in Eq.~(\ref{KG11mar2}) provides the same expression
for the propagator, as derived in the canonical formalism, i.e.%
\begin{equation*}
S_{0}(x-y;\alpha )=i\frac{\delta ^{2}}{\delta \overline{\eta }\delta \eta }
Z_{0}[\eta ,\overline{\eta };\alpha ]|_{\eta \mathbf{=}\overline{\eta }=0}.
\end{equation*}

For interacting fields, we obtain
\begin{equation*}
Z[\overline{\eta },\eta ;\alpha ]=\frac{\exp \left[ i\int dxL_{int}\left(
\frac{1}{i}\frac{\delta }{\delta \overline{\eta }};\frac{1}{i}\frac{\delta }{
\delta \eta }\right) \right] Z_{0}[\eta ,\overline{\eta };\alpha ]}{\exp %
\left[ i\int dxL_{int}\left( \frac{1}{i}\frac{\delta }{\delta \overline{\eta
}};\frac{1}{i}\frac{\delta }{\delta \eta }\right) \right] Z_{0}[\eta ,%
\overline{\eta };\alpha ]|_{\overline{\mathbf{\eta }}=\mathbf{\eta }=0}}.
\end{equation*}%
It is important to note that, when $\alpha \rightarrow \infty $ we have to
recover the flat space-time field theory, for both \ bosons and fermions.

\subsection{Gauge fields}

The Lagrangian density for quantum chromodynamics is given by
\begin{eqnarray*}
\mathcal{L} & = & \overline{\psi }(x)[iD_{\mu }\gamma ^{\mu }-m]\psi (x)-%
\frac{1}{4}F_{\mu \nu }F^{\mu \nu } \\
& &-\,\frac{1}{2\sigma }(\partial ^{\mu }A_{\mu }^{r}(x))^{2}+A_{\mu
}^{r}(x)t^{r}J^{\mu }\left( x\right) +\partial ^{\mu }\chi ^{\ast }(x)D_{\mu
}\chi (y),
\end{eqnarray*}
where
\begin{equation*}
F_{\mu \nu }^{r}=\partial _{\mu }A_{\nu }^{r}(x)-\partial _{\nu }A_{\mu
}^{r}(x)+gc^{rsl}A_{\mu }^{s}(x)A_{\nu }^{l}(x)
\end{equation*}
and $F_{\mu \nu }=\sum\limits_{r}F_{\mu \nu }^{r}t^{r}$ is the field tensor
describing gluons; $t^{r}$ and $c^{rsl}$ are, respectively, generators and
structure constants of the gauge group $SU(3)$; the covariant derivative is
given by $D_{\mu }=\partial _{\mu }+igA_{\mu }=\partial _{\mu }+igA_{\mu
}^{r}(x)t^{r}$ and $\psi (x)$ is the quark field, including the flavor and
color components. The ghost field is given by $\chi (x)$. The quantity $%
\frac{1}{2\sigma }(\partial ^{\mu }A_{\mu }^{r}(x))^{2} $ is the gauge term,
with $\sigma $ being the gauge-fixing parameter.

The generating functional using the Lagrangian density $\mathcal{L}$ is
\begin{eqnarray*}
Z[J,\eta ,\overline{\eta },\xi ,\xi ^{\ast }] & = &\int DAD\psi D\overline{%
\psi }D\chi D\chi ^{\ast } \\
& & \times\,\exp \left[ i\int d^{4}x\left( \mathcal{L}+AJ+\overline{\eta }%
\psi + \overline{\psi }\eta +\xi ^{\ast }\chi +\chi ^{\ast }\xi \right) %
\right] ,
\end{eqnarray*}%
where $\xi ^{\ast }\ $\ and $\xi $ are Grassmann variables describing
sources for ghost fields, and $\overline{\eta }$ and $\eta $ are the
Grassman-variable sources for quarks fields, and $J$ stands for the source
of the gluon-field. Notice that we are using non-tilde fields, in such a way
that the propagator is a c-number.

We write the Lagrangian density in terms of interacting and noninteracting
parts: $\mathcal{L}=\mathcal{L}_{0}+\mathcal{L}_{I}$ with $\mathcal{L}_{0}=%
\mathcal{L}_{0}^{G}+\mathcal{L}_{0}^{FP}+\mathcal{L}_{0}^{Q},$ where $%
\mathcal{L}_{0}^{G}$ is the free gauge field contribution including a gauge fixing term, i.e.
\begin{equation*}
\mathcal{L}_{0}^{G}=-\frac{1}{4}(\partial _{\mu }A_{\nu }^{r}-\partial _{\nu
}A\mu ^{r})(\partial ^{\mu }A^{\nu r}-\partial ^{\nu }A^{\mu r})-\frac{1}{%
2\sigma }(\partial ^{\mu }A_{\mu }^{r})^{2}.
\end{equation*}
The term $\mathcal{L}_{0}^{FP} $ corresponds to the Faddeev-Popov field,
\begin{equation*}
\mathcal{L}_{0}^{FP}=(\partial ^{\mu }\chi _{\mu }^{r\ast })(\partial ^{\mu
}\chi _{\mu }^{r}),
\end{equation*}%
and $\mathcal{L}_{0}^{Q}$ describes the quark field,%
\begin{equation*}
\mathcal{L}_{0}^{F}=\overline{\psi }(x)[\gamma \cdot i\partial -m]\psi (x).
\end{equation*}%
The interaction term is
\begin{eqnarray*}
\mathcal{L}_{I}& = & -\frac{g}{2}c^{rsl}(\partial _{\mu }A_{\nu
}^{r}-\partial _{\nu }A_{\mu }^{r})A^{s\mu }A^{l\nu } \\
& & -\,\frac{g^{2}}{2}c^{rst}c^{ult} A_{\mu }^{r}A_{\nu }^{s}A^{u\mu
}A^{l\nu } \\
& & -\,gc^{rsl}(\partial ^{\mu }\chi ^{r\ast })A_{\mu }^{l}\chi ^{s}(y)+g%
\overline{\psi }t^{r}\gamma ^{\mu }A_{\mu }^{r}\psi .
\end{eqnarray*}

Following steps similar to those in the scalar field case, we write for the
gauge field
\begin{equation*}
Z_{0}^{G(rs)}[J]=\exp \{\frac{i}{2}\int dxdy[J^{\mu }(x)D_{0\mu \nu
}^{(rs)}(x-y;\alpha )J^{\nu }(y),
\end{equation*}%
where
\begin{equation*}
D_{0}^{(rs)\mu \nu }(x;\alpha )= \int \frac{d^{D}k}{(2\pi )^{D}}%
\,e^{-ikx}D_{0}^{(rs)\mu \nu }(k;\alpha )
\end{equation*}%
with
\begin{equation*}
D_{0}^{(rs)\mu \nu }(k;\alpha )=\delta ^{rs}d^{\mu \nu }(k)G_{0}(x-y;\alpha
),
\end{equation*}%
and%
\begin{equation*}
d^{\mu \nu }(k)=g^{\mu \nu }-(1-\sigma )\frac{k^{\mu }k^{\nu }}{k^{2}}.
\end{equation*}

For the Fadeev-Popov field we have
\begin{equation*}
Z_{0}^{FP}[\overline{\xi },\xi ]=\exp \{\frac{i}{2}\int dxdy[\overline{\xi }
(x)G_{0}(x-y;\alpha )\xi (y)] \},
\end{equation*}%
where $\overline{\xi }$ and $\xi $ are Grassmann variables. It is important
to note that $G_{0}(x-y;\alpha )$ is the propagator for the scalar field.
Then we write the full generating functional for the non-abelian gauge field
as
\begin{equation*}
Z[J,\overline{\xi },\xi \mathbf{,}\overline{\eta },\eta ]=\frac{\mathcal{E}
[\partial _{source}]Z_{0}[J,\overline{\xi },\xi \mathbf{,}\overline{\eta }%
,\eta ]}{\mathcal{E}[\partial _{source}]Z_{0}[0]},
\end{equation*}%
where%
\begin{equation*}
\mathcal{E}[\partial _{source}]=\exp \left[ i\int dxL_{int}\left( \frac{1}{i}
\frac{\delta }{\delta J},\frac{1}{i}\frac{\delta }{\delta \overline{\xi }},
\frac{1}{i}\frac{\delta }{\delta \xi },\frac{1}{i}\frac{\delta }{\delta
\overline{\eta }},\frac{1}{i}\frac{\delta }{\delta \eta }\right) \right]
\end{equation*}%
and%
\begin{equation*}
Z_{0}[0]=Z_{0}^{G(rs)}[J]Z_{0}^{FP(rs)}[\overline{\xi },\xi ]Z_{0}^{F(rs)}[%
\overline{\eta },\eta ]|_{J=\overline{\xi }=\xi =\overline{\eta }=\eta =0}.
\end{equation*}

As an example, we derive the gluon-quark-quark three point function to order
$g,$
\begin{eqnarray*}
G_{\mu }^{a}(x_{1},x_{2},x_{3};\alpha )& = & -t^{a}\int \frac{d^{D}p_{1}}{%
(2\pi )^{D}}\frac{d^{D}p_{2}}{(2\pi )^{D}} \\
& & \times\, \exp i\{-p_{1}\cdot x_{1}+p_{2}\cdot x_{2}+(p_{1}-p_{2})\cdot
x_{3}\}d_{\mu \nu } \\
& & \times\, S_{0}(p_{1};\alpha )\gamma ^{v}S_{0}(p_{1};\alpha
)D_{0}(p_{1}-p_{2};\alpha ).
\end{eqnarray*}%
We observe that $G_{\mu }^{a}(x_{1},x_{2},x_{3};\alpha )$ has a part
independent of the topology, i.e. the flat space contribution $G_{\mu
}^{a}(x_{1},x_{2},x_{3})$. This is due to the form of the integral
representation of the propagators $S_{0}(p_{1};\alpha )$ and $%
D_{0}(p_{1}-p_{2};\alpha )$ and represent a general property of the theory.

\section{$S$-Matrix and reaction rates}

In this section we explore the notion of the $S$-matrix, using the GNS
construction as presented earlier. We use the canonical formalism for
abelian fields and derive the reaction rate formulas as functions of
parameters describing the space-time compactification, particularizing our
discussion to the $4$-dimensional Minkowski space.

\subsection{$S$-Matrix}

Consider a field operator $\phi (x)$\thinspace\ such that%
\begin{align*}
\lim_{t\rightarrow -\infty }\phi (x;\alpha ) &=\phi _{in}(x;\alpha ), \\
\lim_{t\rightarrow \infty }\phi (x;\alpha ) &=\phi _{out}(x;\alpha ),
\end{align*}%
where $\phi _{in}(x;\alpha )$ and $\phi _{out}(x;\alpha )$ stand for the in-
and out-fields before and after interaction takes place, respectively. These
two fields are assumed to be related by a canonical transformation
\begin{equation*}
\phi _{out}(x;\alpha )=S^{-1}\phi _{in}(x;\alpha )S,
\end{equation*}%
where $S$ is a unitary operator.

We define the evolution operator, $U(t,t^{\prime }),\,$\ relating the
interacting field to the incoming field, $i.e$
\begin{equation}
\phi (x;\alpha )=U^{-1}(t,-\infty )\phi _{in}(x;\alpha )U(t,-\infty ),
\label{heis14}
\end{equation}%
with $U(-\infty ,-\infty )=1.$ The operator $\phi (x;\alpha )$ satisfies the
Heisenberg equation
\begin{equation*}
-i\partial _{t}\phi (x;\alpha )=[\widehat{H},\phi (x;\alpha )],
\end{equation*}
where the generator of time translation, $\widehat{H}$, is written as $%
\widehat{H}=\widehat{H}_{0}+\widehat{H}_{I}$, with $H_{0}$ and $H_{I}$ being
the free-particle and interaction Hamiltonians, respectively. The field $%
\phi _{in}(x;\alpha )$ satisfies
\begin{equation}
-i\partial _{t}\phi _{in}(x;\alpha )=[\widehat{H}_{0},\phi _{in}(x;\alpha )].
\label{heis24}
\end{equation}

Requiring unitarity of $U(t,t^{\prime }),$ we have
\begin{equation*}
\partial _{t}(U(t,t^{\prime })U^{-1}(t,t^{\prime }))=0.
\end{equation*}%
In addition, from Eq.~(\ref{heis14}) we have
\begin{align*}
\partial _{t}\phi _{in}(x;\alpha ) &=\ \partial _{t}[U(t,-\infty )\phi
(x;\alpha )U^{-1}(t,-\infty )] \\
&=[U(t,-\infty )\partial _{t}U^{-1}(t,-\infty )+i\widehat{H},\phi
_{in}(x;\alpha )].
\end{align*}%
Comparing with Eq.~(\ref{heis24}), we obtain%
\begin{equation*}
i\partial _{t}U(t,-\infty )=\widehat{H}_{I}(t)U(t,-\infty ).
\end{equation*}%
This equation is written as,
\begin{equation*}
U(t,-\infty )=I-i\int_{-\infty }^{t}dt_{1}\widehat{H}_{I}(t_{1})U(t_{1},-%
\infty ),
\end{equation*}%
that is solved by iteration, resulting in%
\begin{eqnarray*}
U(t,-\infty ) & = & I-i\int_{-\infty }^{t}dt_{1}\widehat{H}%
_{I}(t_{1})+(-i)^{2} \int_{-\infty }^{t}\int_{-\infty }^{t_{1}}dt_{1}dt_{2}%
\widehat{H} _{I}(t_{1})\ \widehat{H}_{I}(t_{2})+... \\
& & +\, (-i)^{n}\int_{-\infty }^{t}...\int_{-\infty
}^{t_{n-1}}dt_{1}...dt_{n} \widehat{H}_{I}(t_{1})\ ...\widehat{H}%
_{I}(t_{n})+... \\
& = & \mathrm{T}\exp \left[ -i\int_{-\infty }^{t}dt^{\prime }\widehat{H}
_{I}(t^{\prime })\right] ,
\end{eqnarray*}
where $\mathrm{T}$ is the time-ordering operator.

The $S$-matrix is defined by $S=\lim_{t\rightarrow \infty }U(t,-\infty ),$
such that $S=\sum_{n=0}^{\infty }S^{(n)}$, where%
\begin{equation*}
S^{(n)}=\frac{(-i)^{n}}{n!}\int_{-\infty }^{\infty }...\int_{-\infty
}^{\infty }dt_{1}...dt_{n}T\left[ \widehat{H}_{I}(t_{1})\ ...\widehat{H}%
_{I}(t_{n})\right] .
\end{equation*}%
Then we have
\begin{equation*}
S=\mathrm{T}\exp \left[ -i\int_{-\infty }^{\infty }dt^{\prime }\widehat{H}%
_{I}(t^{\prime })\right] .
\end{equation*}%
The transition operator, $\mathcal{T}$ , is defined by $\mathcal{T}=S-I.$
Observe that $\widehat{H}(\alpha )\equiv \widehat{H},$ and in the definition
of the $S$-matrix there is no need to introduce a tilde $S$-matrix, as is
the case of TFD \cite{kha1}. Here this is a consequence of the GNS
construction.

\subsection{Reaction rates}

Consider the scattering process
\begin{equation*}
p_{1}+p_{2}+...+p_{r}\rightarrow p_{1}^{\prime }+p_{2}^{\prime
}+...+p_{r}^{\prime },
\end{equation*}%
where $p_{i}$ and $p_{i}^{\prime }$ are momenta of the particles in the
initial and final state, respectively. The amplitude for this process is
obtained by the usual Feynman rules as
\begin{equation*}
\left\langle f\right\vert S\left\vert i\right\rangle =\sum_{n=0}^{\infty
}\left\langle f\right\vert S^{(n)}\left\vert i\right\rangle ,
\end{equation*}%
where $\left\vert i\right\rangle =a_{p_{1}}^{\dagger }a_{p_{2}}^{\dagger
}...a_{p_{r}}^{\dagger }\left\vert 0\right\rangle $ and $\left\vert
f\right\rangle =a_{p_{1}^{\prime }}^{\dagger }a_{p_{2}^{\prime }}^{\dagger
}...a_{p_{r}^{\prime }}^{\dagger }\left\vert 0\right\rangle $ with $%
\left\vert 0\right\rangle $ being the vacuum state, such that $%
a_{p}\left\vert 0\right\rangle =0$. For the topology $\Gamma _{D}^{d},$ a
similar procedure may be used by just replacing $\left\vert i\right\rangle $
and $\left\vert f\right\rangle $ states for $\left\vert i;\alpha
\right\rangle $ and $\left\vert f;\alpha \right\rangle $ . The amplitude for
the process is then given as%
\begin{equation*}
\left\langle f;\alpha \right\vert \hat{S}\left\vert i;\alpha \right\rangle
=\sum_{n=0}^{\infty }\left\langle f;\alpha \right\vert \hat{S}^{n}\left\vert
i;\alpha \right\rangle ,
\end{equation*}%
where
\begin{align*}
\left\vert i;\alpha \right\rangle & =a_{p_{1}}^{\dagger }(\alpha
)a_{p_{2}}^{\dagger }(\alpha )...a_{p_{r}}^{\dagger }(\alpha )\left\vert
0(\alpha )\right\rangle , \\
\left\vert f;\alpha \right\rangle & =a_{p_{1}^{\prime }}^{\dagger }(\alpha
)a_{p_{2}^{\prime }}^{\dagger }(\alpha )...a_{p_{r}^{\prime }}^{\dagger
}(\alpha )\left\vert 0(\alpha )\right\rangle ,
\end{align*}%
The vacuum state in the topology $\Gamma _{D}^{d}$ is given by $\left\vert
0(\alpha )\right\rangle $. As emphasized earlier, the phase-space factors
are not changed by the topology. The meaning of these states is described in
Section IV.

The differential cross-section for the particular process
\begin{equation*}
p_{1}+p_{2}\rightarrow p_{1}^{\prime }+p_{2}^{\prime }+...+p_{r}^{\prime }
\end{equation*}
is given by
\begin{eqnarray}
d\sigma &=&(2\pi )^{4}\delta ^{4}(p_{1}^{\prime }+p_{2}^{\prime
}+p_{3}^{\prime }+...+p_{r}^{\prime }-p_{1}-p_{2})  \notag \\
& &\times\, \frac{1}{4E_{1}E_{2}v_{rel}}\prod_{l}(2m_{l})\prod_{j=1}^{r}%
\frac{ d^{3}p_{j}^{\prime }}{(2\pi )^{3}2E_{j}^{\prime }}\left\vert
M_{fi}(\alpha )\right\vert ^{2},  \label{5a}
\end{eqnarray}%
where $E_{j}^{\prime}=\sqrt{m_{j}^{^{\prime}2}+\mathbf{p}_{j}^{^{\prime}2}}$
and $v_{rel}$ is the relative velocity of the two initial particles with
3-momenta $\mathbf{p}_{1}$ and\textbf{\ }$\mathbf{p}_{2}$. The factor $%
2m_{j} $ appears for each lepton in the initial and final state. Here $E_{1}$
and $E_{2}$ are the energies of the two particles with momenta $\mathbf{p}%
_{1}$ and\textbf{\ }$\mathbf{p}_{2}$, respectively. The amplitude $M_{fi}$
is related to the $S$-matrix element by
\begin{equation}
\left\langle f;\alpha \right\vert S\left\vert i;\alpha \right\rangle
=i\left( 2\pi \right) ^{4}M_{fi}(\alpha )\prod_{i}\left( \frac{1}{2VE_{i}}%
\right) ^{\frac{1}{2}}\prod_{f}\left[ \frac{1}{2VE_{f}}\right] ^{\frac{1}{2}
}\delta ^{4}(p_{f}-p_{i}).  \label{6}
\end{equation}%
Here $p_{f}$ and $p_{i}$ are the total $4$-momenta in the final and initial
state, respectively. The product extends over all the external fermions and
bosons, with $E_{i}$ and $E_{f}$ being the energy of particles in the
initial and final states, respectively and $V$ is the volume.

\subsection{Decay of particles}

Consider the decay of the boson field $\sigma $ into $\pi $ , with an
interaction Lagrangian density given by
\begin{equation}
\mathcal{L}_{I}=\lambda \sigma (x)\pi \,(x)\pi (x).  \label{7}
\end{equation}%
The initial and final states in $\Gamma _{D}^{d}$ are, respectively,
\begin{equation*}
\left\vert i;\alpha \right\rangle =a_{k}^{\dag }(\alpha )\left\vert 0(\alpha
)\right\rangle ,
\end{equation*}%
and
\begin{equation*}
\left\vert f;\alpha \right\rangle =b_{k_{1}}^{\dag }(\alpha )b_{k_{2}}^{\dag
}(\alpha )\left\vert 0(\alpha )\right\rangle ,
\end{equation*}%
where $a_{k}^{\dag }(\alpha )$ and $b_{k}^{\dag }(\alpha )$ are creation
operators in the topology $\Gamma _{D}^{d}$ for the $\sigma $- and $\pi $-
particles, respectively, with momenta $k$. At the tree level, the transition
matrix element is
\begin{eqnarray*}
\left\langle f;\alpha \right\vert \hat{S}\left\vert i;\alpha \right\rangle &
= & i\lambda \int dx\left\langle 0(\alpha )\right\vert b_{k_{2}}(\alpha
)b_{k_{1}}(\alpha ) \\
& & \times\, \lbrack \sigma (x)\pi (x)\pi (x)-\tilde{\sigma}(x)\tilde{\pi}(x)%
\tilde{\pi}(x)]a_{k}^{\dag }(\alpha )\left\vert 0(\alpha )\right\rangle .
\end{eqnarray*}%
Using the expansion of the boson fields, $\sigma (x)$ and $\pi (x)$, in
momentum space, the Bogoliubov transformation and the commutation relations,
the two terms of the matrix element are calculated. For instance we have
\begin{equation*}
\left\langle 0(\alpha )\right\vert \sigma (x)a_{k}^{\dag }(\alpha
)\left\vert 0(\alpha )\right\rangle =e^{-ikx}\cosh \theta (k;\alpha ),
\end{equation*}%
Combining these factors, the amplitude for the process is
\begin{eqnarray*}
M_{fi}(\beta )& = & \lambda \lbrack \cosh (k;\alpha )\cosh \theta
(k_{1};\alpha )\cosh \theta (k_{2};\alpha ) \\
& & -\, \sinh \theta (k;\alpha )\sinh \theta (k_{1};\alpha )\sinh \theta
(k_{2};\alpha )].
\end{eqnarray*}%
Note that the indices \textquotedblleft 1\textquotedblright\ and
\textquotedblleft 2\textquotedblright\ in $k_{1}$ and $k_{2}$ are refereing
here to particles.

The decay rate for the $\sigma $-meson is given as
\begin{eqnarray}
\Gamma (w,\alpha ) &=&\frac{1}{2w}\int \frac{d^{3}k_{1}d^{3}k_{2}(2\pi
)^{4}\delta ^{4}(k-k_{1}-k_{2})}{(2w_{1})(2w_{2})(2\pi )^{3}(2\pi )^{3}}%
\left\vert M_{fi}(\alpha )\right\vert ^{2}  \notag \\
&=&\frac{\lambda ^{2}}{32w\pi ^{2}}\int \frac{d^{3}k_{1}}{w_{1}}\frac{%
d^{3}k_{2}}{w_{2}}\delta ^{4}(k-k_{1}-k_{2})W(w;w_{1},w_{2};\alpha ),
\label{May24101}
\end{eqnarray}
where%
\begin{eqnarray*}
W(w;w_{1},w_{2};\alpha )& = & |\cosh \theta (k;\alpha )\cosh \theta
(k_{1};\alpha )\cosh \theta (k_{2};\alpha ) \\
& & -\, \sinh \theta (k;\alpha )\sinh \theta (k_{1};\alpha )\sinh \theta
(k_{2};\alpha )|^{2},
\end{eqnarray*}
with
\begin{equation*}
w_{i}=\sqrt{\kappa _{i}^{2}+m^{2}},w=\sqrt{\mathbf{k}^{2}+M^{2}}.
\end{equation*}%
Using $\sinh ^{2}\theta (k;\alpha )=v^{2}(k_{\alpha };\alpha )\equiv
n(k;\alpha )$ and $\cosh ^{2}\theta (k;\alpha )=u^{2}(k;\alpha )\equiv
1+n(k;\alpha )$, we have%
\begin{eqnarray}
W(w;w_{1},w_{2};\alpha ) &=&\left\vert \sqrt{1+n(k;\alpha )}\right\vert
^{2}\left\vert \sqrt{1+n(k_{1};\alpha )}\right\vert ^{2}\left\vert \sqrt{%
1+n(k_{2};\alpha )}\right\vert ^{2}  \notag \\
&&+\left\vert \sqrt{n(k;\alpha )}\right\vert ^{2}\left\vert \sqrt{%
n(k_{1};\alpha )}\right\vert ^{2}\left\vert \sqrt{n(k_{2};\alpha )}%
\right\vert ^{2}  \notag \\
&&-\,2\,\mathfrak{Re}\{[1+n(k;\alpha )][1+n(k_{1};\alpha )][1+n(k_{2};\alpha
)]  \notag \\
&&\times\, n(k;\alpha )n(k_{1};\alpha )n(k_{2};\alpha )\}^{1/2}.
\end{eqnarray}

Considering the rest frame of the decaying particle: $w=M,$ $\mathbf{k}%
=0,\,w_{i}=\sqrt{\mathbf{k}_{i}^{2}+m^{2}}=\sqrt{\mathbf{q}^{2}+m^{2}}
=w_{q} $ , and the case of temperature only, i.e. $\alpha =(\beta ,0,0,0)$,
we recover the result derived in Ref.~\cite{khaabd1}.

One aspect to be emphasized is the notion of quasi-particles. The energy
spectrum of the particles taking place in the reaction has changed as a
consequence of the compactification. This new spectrum corresponds to the
energy of the quasi-particles. The broken symmetry here is due to a
topological specification in the Minkowski space-time. This interpretation
is valid also for the thermal effects, considered from a topological point
of view.

\section{Concluding remarks}

In this paper we have developed a theory for quantum fields defined on a $D$
-dimensional space-time having a topology $\Gamma _{D}^{d}=(\mathbb{S}%
^{1})^{d}\times \mathbb{M}^{D-d}$, with $1\leq d\leq D$. This describes
simultaneously spatial constraints and thermal effects. We use the modular
construction of c$^{\ast }$-algebra as a key tool. With the modular group,
we study a $^{\ast }$-representation of Lie algebras, specifically analyzing
the Poincar\'{e} group. The propagator for bosons and fermions is found to
be a generalization of the Fourier integral representation~\cite{baym11,3dj}
of the imaginary-time propagator. Some results deserve to be emphasized.

\begin{itemize}
\item[(i)] Considering $d=1$, with the compactification parameter being $%
\beta =1/T$, we show that it is possible to develop a consistent real-time
quantum field theory at finite temperature $T$, where the propagator is
given in the Fourier integral representation. Therefore, there is no need to
use a $2\times 2$ matrix structure for the propagator and for the generating
functional, as is the case in TFD and in the Schwinger-Keldysh approach.
This is a simplification in the formalism, that is explored here for the
general case of the topology $\Gamma _{D}^{d}$.

\item[(ii)] Using the modular representation, we study the Poincar\'{e}
group. The modular conjugation is defined in order to respect the Lie
algebra structure. This procedure provides a precise way to define the
modular conjugation for fermions~\cite{ojima1}.

\item[(iii)] The extension of the formalism for abelian and non-abelian
gauge-fields is developed using functional methods.

\item[(iv)] The S-matrix is introduced, and as an application we calculate
the space-compactification effect in a decay process.

\item[(v)] The compactification is described as a process of condensation in
the vacuum state, $|0(\alpha )\rangle $. The parameter $\alpha $ describes
the topological effects, which modify the energy spectrum, giving rise to
the notion of quasi-particles.

\item[(vi)] The Fourier integral representation of the propagator is
separated into a divergent part and a finite contribution from the
topological effects. This feature has proved to be useful in the study of numerous
processes~\cite{kha1,3belac1}.
\end{itemize}

\textbf{Acknowledgements}: The authors thank the agencies NSERC from Canada
and CAPES, CNPq and FAPERJ, from Brazil, for financial support.

\appendix

\section{Fourier integral representation of the Green function}

Consider a scalar field at finite temperature, such that the Green function
satisfies the Klein-Gordon equation written in a $D$-dimensional Minkowski
space,
\begin{equation}
(\square +m^{2})G(x-y;\beta )=-\delta (x-y).  \label{bo1}
\end{equation}%
We take the Fourier integral representation of the Green function as derived
in Refs.~\cite{baym11} and \cite{3dj}, written for $\mathbb{M}^{D}$,
\begin{equation}
G_{0}(x-y;\beta )=\int \ \frac{{d^{D}p}}{(2\pi )^{D}}{\ G_{0}(p;\beta )},
\end{equation}%
where%
\begin{equation*}
G_{0}(p;\beta )=G_{0}(p)+f_{\beta }(p^{0})[G_{0}(p)-G_{0}^{\ast }(p)] ,
\end{equation*}%
where
\begin{equation*}
f_{\beta}(p^0) = \frac{1}{e^{\beta p^0} - 1} = \sum_{l_0 = 1}^{\infty} e^{- \beta p^0 l_0}
\end{equation*}
and
\begin{equation*}
G_{0}(k)=\frac{-1}{p^{2}-m^{2}+i\varepsilon }.
\end{equation*}%
The thermal theory corresponds to a topology $\mathbb{S}^{1}\times \mathbb{M}%
^{D-1}$. Now we would like to generalize the theory to the topology $\mathbb{%
S}^{1}\times \mathbb{S}^{1}\times \mathbb{M}^{D-2}$, $i.e.$,
compactification of time and one spatial dimension. We choose that the $x^{1}
$ direction is compactified with a period $L_{1}$. Observe that the topology
does not change the local properties of the system. This implies that
locally the Minkowski space, as well as a differential equation defined by
an isometry, such as the Klein-Gordon equation, are the same. However, the
topology imposes modifications on boundary conditions to be fulfilled by the
field and the respective Green function.

The new Green function satisfies the periodic boundary condition,
\begin{equation}
G(x^{0},x^{1},\mathbf{x};\beta ) \equiv G(x^{0},x^{1}+L_{1},\mathbf{x};\beta
) = G(x+L_{1}\widehat{{n}}_{1};\beta ),  \label{bo2}
\end{equation}%
where $\widehat{{n}}_{1}=({n}_{1}^{\mu })=(0,1,0,\dots,0)$ and $\mathbf{x} =
(x^2,\dots,x^{D-1})$. A solution of Eq.~(\ref{bo1}), satisfying this
condition, is obtained by using the Fourier expansion
\begin{equation}
G(x-y;L_{1};\beta )=\frac{1}{L_{1}}\sum\limits_{n=-\infty }^{\infty } \int
\frac{dp_{0}d^{D-2}\mathbf{p}}{(2\pi )^{D-1}} \,
e^{-ip_{n}(x-y)}G_{0}(p_{n};L_{1};\beta ),  \label{bogg1}
\end{equation}%
where%
\begin{equation*}
p_{n}=(p_{0},p_{1n},\mathbf{p}),\ \ p_{1n}=\frac{2\pi n}{L_{1}},\ \ \mathbf{p%
} = (p_2,\dots,p_{D-1}),
\end{equation*}%
\begin{equation}
G_{0}(p_{n};L_{1};\beta )=G_{0}(p_{n};L_{1}) + f_{\beta }(p^{0}) [G_{0}(p_{n};L_{1})-G_{0}^{\ast }(p_{n};L_{1})]  \label{bogapr21}
\end{equation}%
and
\begin{equation}
G_{0}(p_{n};L_{1})=\frac{-1}{p_{n}^{2}-m^{2}+i\varepsilon}.  \label{boggjul1}
\end{equation}%
Inversely, we have,%
\begin{equation}
G(p_{n};L_{1};\beta )=\int_{0}^{L_{1}}dx^{1}\int dx^{0}d^{D-2}\mathbf{x}%
\,e^{ip_{n} x}G(x;L_{1};\beta ).  \label{dj0}
\end{equation}

To obtain the Fourier integral representation of $G(x-y;L_{1};\beta )$, we first perform
a sort of Wick rotation such that $L_1 \rightarrow - i L'_1$; this allows us to proceed in a similar faction
as in the case of temperature. Then, we write $G(x-y;L'_{1};\beta )$ as
\begin{equation}
G(x-y;L'_{1};\beta )=\theta (x^{1}-y^{1})G^{>}(x-y;L'_{1};\beta )+\theta
(y^{1}-x^{1})G^{<}(x-y;L'_{1};\beta )  \label{dj11}
\end{equation}%
and, from Eq.~(\ref{bo2}), we have%
\begin{equation}
G^{<}(x;L'_{1};\beta )\left. {}\right\vert _{x^{1}=0}=G^{>}(x;L'_{1};\beta
)\left. {}\right\vert _{x^{1}=-i L'_{1}}.  \label{dj110}
\end{equation}%
Then, Eq.~(\ref{dj0}) reads%
\begin{equation}
G(p'_{n};L'_{1};\beta )=\int_{0}^{-i L'_{1}}dx^{1}\int dx^{0}d^{D-2}\mathbf{x}%
\,e^{ip'_{n}x}G^{>}(x;L'_{1};\beta ) ,  \label{dj111}
\end{equation}%
where $p'_{1 n} = 2\pi n/(-i L'_1)$.
The Fourier integral transform of $G(x-y;L'_{1};\beta ),$ denoted by $%
\overline{G}(p;L'_{1};\beta ),$ is%
\begin{equation}
\overline{G}(p;L'_{1};\beta )=\overline{G}^{(1)}(p;L'_{1};\beta )+\overline{G}
^{(2)}(p;L'_{1};\beta ),  \label{dj112}
\end{equation}%
where%
\begin{align}
\overline{G}^{(1)}(p;L'_{1};\beta )& =\int d^{D}x\,e^{ipx}\theta
(x^{1})G^{>}(x;L'_{1};\beta ),  \label{dj12} \\
\overline{G}^{(2)}(p;L'_{1};\beta )& =\int d^{D}x\,e^{ipx}\theta
(-x^{1})G^{<}(x;L'_{1};\beta ).  \label{dj13}
\end{align}%
Writing
\begin{equation}
G^{>}(x;L'_{1};\beta )=\int \frac{d^{D}p}{(2\pi )^{D}}\,e^{-ipx}\overline{G}%
^{>}(p;L'_{1};\beta )  \label{dj160}
\end{equation}%
and using the integral representation of the step function,%
\begin{equation*}
\int dk^{1}\frac{e^{-ik^{1}x^{1}}}{k^{1}+p^{1}+i\varepsilon }=(-2\pi
i)e^{ip^{1}x^{1}}\theta (x^{1}),
\end{equation*}%
in Eq.~(\ref{dj12}), we have%
\begin{equation}
\overline{G}^{(1)}(p;L'_{1};\beta )=i\int \frac{dk^{1}}{2\pi }\frac{\overline{
G}^{>}(p_{0},k_{1},\mathbf{p};L'_{1};\beta )}{k^{1}-p^{1}+i\varepsilon }.
\label{dj16}
\end{equation}%
With
\begin{equation*}
G^{<}(x;L'_{1};\beta )=\int \frac{d^{D}p}{(2\pi )^{D}}e^{-ipx}\overline{G}%
^{<}(p;L'_{1};\beta ),
\end{equation*}%
and using the integral representation of the step function,%
\begin{equation*}
\int dk^{1}\frac{e^{-ik^{1}x^{1}}}{k^{1}+p^{1}-i\varepsilon }=2\pi
ie^{ip^{1}x^{1}}\theta (-x^{1}),
\end{equation*}%
in Eq.~(\ref{dj13}), we obtain%
\begin{equation}
\overline{G}^{(2)}(p;L'_{1};\beta )=-i\int \frac{dk^{1}}{2\pi }\frac{%
\overline{G}^{<}(p_{0},k_{1},\mathbf{p};L'_{1};\beta )}{k^{1}-p^{1}-i%
\varepsilon }.  \label{dj17}
\end{equation}%
Substituting Eqs.~(\ref{dj16}) and (\ref{dj17}) in Eq.~(\ref{dj112}), we get
\begin{equation}
\overline{G}(p;L'_{1};\beta )=i\int \frac{dk^{1}}{2\pi }\left[ \frac{%
\overline{G}^{>}(p_{0},k_{1},\mathbf{p};L'_{1};\beta )}{k^{1}-p^{1}+i%
\varepsilon }-\frac{\overline{G}^{<}(p_{0},k_{1},\mathbf{p};L'_{1};\beta )}{
k^{1}-p^{1}-i\varepsilon }\right]  \label{dj171}
\end{equation}

>From the periodicity of the Green function, we have%
\begin{equation}
\overline{G}^{<}(p;L'_{1};\beta )=e^{L'_{1} p^{1}}\overline{G}%
^{>}(p;L'_{1};\beta ).  \label{dj18101}
\end{equation}%
Defining%
\begin{equation*}
f_{L'_{1}}(p^{1})=\frac{1}{e^{L'_{1}p^{1}}-1} = \sum_{l_1 = 1}^{\infty} e^{-L'_{1} p^{1} l_1} ,
\end{equation*}%
we write%
\begin{align}
\overline{G}^{>}(p;L'_{1};\beta )& =f_{L'_{1}}(p^{1})A(p;L'_{1};\beta )
\label{dj18} \\
\overline{G}^{<}(p;L'_{1};\beta )& =[f_{L'_{1}}(p^{1})+1]A(p;L'_{1};\beta ).
\label{dj19}
\end{align}%
Then we have
\begin{equation*}
A(p;L'_{1};\beta )=\overline{G}^{<}(p;L'_{1};\beta )-\overline{G}%
^{>}(p;L'_{1};\beta ).
\end{equation*}%
With these results, Eq.~(\ref{dj171}) reads%
\begin{eqnarray}
\overline{G}(p;L'_{1},\beta )& = & i\int \frac{dk^{1}}{2\pi }\left[ \frac{%
f_{L'_{1}}(k^{1})A(p_{0},k_{1},\mathbf{p};L'_{1};\beta )}{k^{1}-p^{1}+i%
\varepsilon }\right.  \notag \\
& & -\,\left. \frac{[f_{L'_{1}}(k^{1})+1]A(p_{0},k_{1},\mathbf{p};L'_{1};\beta
)}{ k^{1}-p^{1}-i\varepsilon }\right] .  \label{dj191}
\end{eqnarray}

We do not have as yet an explicit expression for $A(p;L'_{1};\beta ).$ To
determine this function, we use the fact that we know $G(p'_{n};L'_{1};\beta )$%
, as given in Eq.~(\ref{dj111}). Using Eq.~(\ref{dj160}), we have
\begin{equation*}
G(p'_{n};L'_{1};\beta )=\int_{0}^{-i L'_{1}}dx^{1}\int dx^{0}d^{D-2}\mathbf{x}%
\,e^{ip'_{n}x}\int \frac{d^{D}k}{(2\pi )^{D}}e^{-ikx} \overline{G}%
^{>}(k;L'_{1};\beta ).
\end{equation*}%
>From Eq.~(\ref{dj18}) and the integral representation
\begin{equation*}
\int_{0}^{-i L'_{1}}dx^{1}e^{-i(p_{n}^{\prime 1}-k^{1})x^{1}} =\frac{1}{f_{L'_{1}}(k^{1})%
} \frac{i}{p_{n}^{\prime 1}-k^{1}},
\end{equation*}%
we obtain
\begin{equation*}
G(p'_{n};L'_{1};\beta )=i\int \frac{dk^{1}}{2\pi }\frac{A(p_{0},k_{1},\mathbf{p%
};L'_{1};\beta )}{p_{n}^{\prime 1}-k^{1}},
\end{equation*}%
where $A(p;L'_{1};\beta )$ is the generalization of the spectral function
associated with the momentum $p^{1}$.

We consider the analytic continuation of $G(p'_{n};L'_{1};\beta )$ to take $%
p_{n}^{\prime 1}$ to be a continuum variable, $p^{1}.$ The only possible analytical
continuation of $G(p'_{n};L'_{1};\beta )$ without essential singularity at $%
p\rightarrow \infty $ is the function
\begin{equation*}
\mathcal{G}_{0}(p;L'_{1};\beta )=i\int \frac{dk^{1}}{2\pi }\frac{ A(p_{0},k_{1},%
\mathbf{p};L'_{1};\beta )}{p^{1}-k^{1}},
\end{equation*}%
where, by definition,
\begin{equation}
\mathcal{G}_{0}(p;L'_{1};\beta )=G_{0}(p;L'_{1};\beta ).  \label{dj20}
\end{equation}%
Using this result, we calculate $A(p)$ by showing that
\begin{align*}
\mathcal{G}(p;\varepsilon )& =\mathcal{G}_{0}(p_{0},p^{1}+i\varepsilon ,
\mathbf{p};L'_{1};\beta)-\mathcal{G}_{0}(p_{0},p^{1}-i\varepsilon ,\mathbf{p};
L'_{1};\beta) \\
& =i\int \frac{dk^{1}}{2\pi }A(p_{0},k_{1},\mathbf{p};L'_{1};\beta )\left[
\frac{1 }{p^{1}-k^{1}+i\varepsilon }-\frac{1}{p^{1}-k^{1}-i\varepsilon }%
\right] \\
& =i\int \frac{dk^{1}}{2\pi }A(p_{0},k_{1},\mathbf{p};L'_{1};\beta )(-2\pi
i)\delta (p^{1}-k^{1}).
\end{align*}%
This leads to
\begin{equation*}
A(p;L'_{1};\beta )=\mathcal{G}_{0}(p_{0},p^{1}+i\varepsilon ,\mathbf{p}%
;L'_{1};\beta )-\mathcal{G}_{0}(p_{0},p^{1}-i\varepsilon ,\mathbf{p};L'_{1};\beta )
\end{equation*}%
which describes a discontinuity of $\mathcal{G}_{0}(p;L_1;\beta )$ across
the real axis $p^{1}.$

Using the identity
\begin{equation*}
\delta (x^{2}-y^{2})=\frac{1}{2|y|}\left[ \delta (x+y)+\delta (x-y)\right] ,
\end{equation*}%
the Fourier integral representation of $G(x-y;L'_{1};\beta)$ can be written, after some
calculations and transforming back $L'_1 \rightarrow i L_1$, as
\begin{equation}
G(x-y;L_{1};\beta )=\int \frac{d^{D}p}{(2\pi )^{D}}e^{-ip(x-y)}%
\{G_{0}(p)+f_{L_{1}\beta }(p^{0},p^{1})[G_{0}(p)-G_{0}^{\ast }(p)]\},
\label{bo4}
\end{equation}%
where
\begin{equation}
f_{L_{1}\beta }(p^{0},p^{1})=f_{\beta }(p^{0})+f_{L_{1}}(p^{1})+2f_{\beta
}(p^{0})f_{L_{1}}(p^{1}), \label{fbetaL}
\end{equation}%
with%
\begin{equation*}
f_{L_{1}}(p^{1})=\sum\limits_{l_1=1}^{\infty }e^{-i L_{1} p^{1} l_1}.
\end{equation*}%
In this representation, one important result is that the content of the flat
space is given in a separated term involving only $G_{0}(p)$, while the
topological effect of $\Gamma _{D}^{2}$ is present in the term with $%
f_{L_{1}\beta }(p^{0},p^{1})$, describing compactification of space and
time. In addition we obtain: for $L\rightarrow \infty \,$, $f_{L_{1}\beta
}(p^{0},p^{1})\rightarrow f_{\beta }(p^{0})$, and for $\beta \rightarrow
\infty \,$, $f_{L_{1}\beta }(p^{0},p^{1})\rightarrow f_{L_{1}}(p^{1})$, a
consistent result.

For fermions, due to the nature of the statistics, one obtains the same expression as in
Eq.~(\ref{fbetaL}) but with
\begin{equation*}
f_{\beta}(p^{0})=\sum\limits_{l_0=1}^{\infty } (-1)^{1+l_0} e^{-\beta p^{0} l_0} , \;\;  f_{L_{1}}(p^{1})=\sum\limits_{l=1}^{\infty } (-1)^{1+l_1}  e^{-i L_{1} p^{1} l_1} .
\end{equation*}

The same procedure can be repeated for compactification of several
dimensions, corresponding to the topology $\Gamma _{D}^{d}$, for both boson
and fermion fields. For $d$ ($\leq D$) compactified dimensions, $x^0,x^1,\dots,x^{d-1}$,
the result is the propagator shown in Eqs.~(\ref{bo10121},\ref{CBPF22.12.3}), for bosons, and in
Eqs.~(\ref{Sfermions},\ref{Gxfermi},\ref{Gkfermi}), for fermions,
with $f_{\alpha }(p^{\alpha })$ given by
\begin{eqnarray}
f_{\alpha }(p^{\alpha }) & = & \sum_{s=1}^{d} \sum_{\{\sigma _{s}\}} 2^{s-1}   \left(
\prod_{j=1}^{s} f_{\alpha _{\sigma _{j}}}(p^{\sigma _{j}})\right)   \notag \\
& = & \sum_{s=1}^{d} \sum_{\{\sigma _{s}\}} 2^{s-1} \sum_{l_{\sigma _{1}},...,l_{\sigma _{s}}=1}^{\infty }
(-\eta)^{s+\sum_{r=1}^{s}l_{\sigma _{r}}}\,\exp \{-\sum_{j=1}^{s}\alpha _{\sigma
_{j}}l_{\sigma _{j}}p^{\sigma _{j}}\},  \label{v2gen}
\end{eqnarray}%
where $\eta =1\;(-1)$ for fermions (bosons) and $\{\sigma _{s}\}$ denotes the set
of all combinations with $s$ elements, $\{\sigma _{1},\sigma _{2},...\sigma
_{s}\}$, of the first $d$ natural numbers $\{0,1,2,...,d-1\}$, that is all
subsets containing $s$ elements; in order to obtain the physical condition of finite temperature and
spatial confinement, $\alpha _{0}$ has to be taken as a positive real number, $\beta = T^{-1}$,
while $\alpha _{n}$, for $n=1,2,...,d-1$, must be pure imaginary of the form $%
iL_{n}$.


\begin{thebibliography}{99}
\bibitem{3mats1} T. Matsubara, Prog. Theor. Phys. 14 (1955) 351.

\bibitem{3ume4} H. Ezawa, Y. Tomozawa, H. Umezawa, N. Cimento Ser. X 5
(1957) 810.

\bibitem{3kap1} J.I. Kapusta, Finite-Temperature Field Theory, Cambridge
University Press, Cambridge, 1989.

\bibitem{baym11} L.P. Kadanoff, G. Baym, Quantum Statistical Mechanics,
Benjamin, N. York, 1962.

\bibitem{3dj} L. Dolan, R. Jackiw, Phys. Rev. D 9 (1974) 3320.

\bibitem{birrel1} N.D. Birrell, L.H. Ford, Phys. Rev. D 22 (1980) 330.

\bibitem{kha1} F.C. Khanna, A.P.C Malbouisson, J.M.C. Malbouisson, A.E.
Santana, Thermal Quantum Field Theory: Algebraic Aspects and Applications,
World Scientific, Singapore, 2009.

\bibitem{3kel1} J. Schwinger, J. Math. Phys. 2 (1961) 407.

\bibitem{3kel2} P.M. Bakshi, K.T. Mahanthappa, J. Math. Phys. 4 (1963) 1; 12.

\bibitem{3kel3} K.T. Mahanthappa, Phys. Rev. 126 (1962) 329.

\bibitem{3kel4} L.V. Keldysh, Zh. \'{E}ksp. Toer. Fiz, 47(1964) 1515 [Sov.
Phys. JETP 20 (1965) 1018].

\bibitem{3belac1} M. Le Bellac, Thermal Field Theory, Cambridge University
Press, Cambridge, 1996.

\bibitem{3ume1} Y. Takahashi, H. Umezawa, Coll. Phenomena 2(1975) 55
(Reprinted in Int. J. Mod. Phys. 10 (1996) 1755).

\bibitem{3ume2} H. Umezawa, Advanced Field Theory: Micro, Macro and Thermal
Physics, AIP, New York, 1993.

\bibitem{3das} A. Das, Finite Temperature Field Theory, World Scientific,
Singapore, 1997.

\bibitem{chu1} H. Chu, H. Umezawa, Int. J. Mod. Phys. A 9 (1994) 2363.

\bibitem{birrel2} L.H. Ford, N.F. Svaiter, Phys. Rev D 51 (1995) 6981.

\bibitem{3comp1} A.P.C. Malbouisson, J.M.C. Malbouisson, J. Phys. A: Math.
Gen. 35 (2002) 2263.

\bibitem{3comp2} A.P.C. Malbouisson, J.M.C. Malbouisson, A.E. Santana, Nucl.
Phys. B 631 (2002) 83.

\bibitem{3comp3} J.C. da Silva, F.C. Khanna, A. Matos Neto, A.E. Santana,
Phys. Rev. A 66 (2002) 052101.

\bibitem{3comp4} H. Queiroz, J.C. da Silva, F.C. Khanna, J.M.C Malbouisson,
M. Revzen, A.E. Santana, Ann. Phys. 317 (2005) 220; Erratum and addendum,
Ann. Phys. 321 (2006) 1274.

\bibitem{3comp5} L.M. Abreu, C. de Calan, A.P.C. Malbouisson, J.M.C.
Malbouisson, A.E. Santana, J. Math. Phys. 46 (2005) 12304.

\bibitem{3comp6} A.P.C. Malbouisson, J.M.C. Malbouisson, A.E. Santana, J.C.
da Silva, Phys. Lett. B 583 (2004) 373.

\bibitem{KMMS10} F.C. Khanna, A.P.C Malbouisson, J.M.C. Malbouisson, A.E.
Santana, EuroPhys. Lett. 92 (2010) 11001.

\bibitem{ford3} L.H. Ford, T. Yoshimura, Phys. Lett. A 70 (1979) 89.

\bibitem{genr8} L.H. Ford, Phys. Rev. D 21 (1980) 933.

\bibitem{chodos22} A. Chodos, E. Myers, Ann. Phys. 156 (1984) 412.

\bibitem{tori1} N.L. Balaz, A. Voros, Phys. Rep. 143 (1986) 109.

\bibitem{eliz1} E. Elizalde, K. Kirsten, J. Math. Phys. 35 (1994) 1260.

\bibitem{gen6} H. Kleinert, A. Zhuk, Theor. Math. Phys. (Russ) 108 (1996)
482.

\bibitem{ito1} M. Ito, Nucl. Phys. B 668 (2003) 322.

\bibitem{gen3} Yu. P. Goncharov, Russ. Phys. J. 26 (2004) 752.

\bibitem{genr9} N. Ahmadi, M. Nouri-Zonoz, Phys. Rev. D 71 (2005) 104012.

\bibitem{gen7} J.L. Tomazelli, L.C. Costa, Int. J. Theor. Phys. 45 (2006)
499.

\bibitem{gen1} N. Ahmadi, M. Nouri-Zonoz, Nucl. Phys. B 738 (2006) 269.

\bibitem{luc1} L.M. Abreu, M. Gomes, A.J. da Silva, Phys. Lett. B 642 (2006)
551.

\bibitem{luc2} L.M. Abreu, A.P.C. Malbouisson, J.M.C. Malbouisson, A.E. Santana, Nucl.
Phys. B 819 (2009) 127.

\bibitem{luc3} L.M. Abreu, A.P.C. Malbouisson, J.M.C. Malbouisson, EuroPhys. Lett. 90
(2010) 11001.

\bibitem{mul1} M.P. Lima, D. M\"uller, Class. Quantum Grav. 24 (2007) 897.

\bibitem{genr4} A.A. Saharian, Class. Quantum Grav. 25 (2008) 165012.

\bibitem{ebe1} D. Ebert, K.G. Klimenko, A.V. Tyukov, V.Ch. Zhukovsky, Phys.
Rev D 78 (2008) 045008.

\bibitem{ebe2} S. Bellucci, A.A. Saharian, Phys. Rev. D 80 (2009) 105003.

\bibitem{ebe5} A.A. Saharian, A.L. Mkhitaryan, Eur. Phys. J. C 66 (2010) 295.

\bibitem{aco1} A. Connes, Noncommutative geometry, Springer, Berlin, 1992.

\bibitem{aco2} A. Connes, M.R. Douglas, A. Schwarz, J. High Energy Phys. 2
(1998) 3.

\bibitem{ebe3} T. Prosen, Phys. Rev. E 60 (1999) 1658.

\bibitem{kraj1} T. Krajewski, R. Wulkenhaar, Int. J. Mod. Phys. A 15 (2000)
1011.

\bibitem{emch} G.G. Emch, Algebraic Methods in Statistical Mechanics and
Quantum Field Theory, Wiley-Interscience, New York, 1972.

\bibitem{brate} O. Bratteli, D.W. Robinson, Operator Algebras and Quantum
Statistical Mechanics, vols. I and II, Springer, Berlin, 1979.

\bibitem{tomita} M. Takesaki, Tomita Theory of Modular Hilbert Algebras and
its Applications, Springer-Verlag, Berlin, 1970.

\bibitem{ojima1} I. Ojima, Ann. Phys. 137 (1981) 1.

\bibitem{ade31} A.E. Santana, A. Matos Neto, J.D.M. Vianna, F.C. Khanna,
Int. J. Theor. Phys. 38 (1999) 641.

\bibitem{ade44} A.E. Santana, A. Matos Neto, J.D.M. Vianna, F.C. Khanna,
Physica A 280 (2000) 405.

\bibitem{kha1000} F.C. Khanna, A.P.C Malbouisson, J.M.C. Malbouisson, A.E.
Santana, Ann. Phys. 324 (2009) 1931.

\bibitem{ade1} A.E. Santana, F.C. Khanna, Phys. Lett. A 203 (1995) 68.

\bibitem{khaabd1} A.M. Rakhimov, F.C. Khanna, Phys. Rev. C 64 (2001) 064907.
\end{thebibliography}
\end{document}